\documentclass[12pt]{article}

\usepackage{latexsym,amssymb,amsgen,amsmath,amsxtra, epsfig,
amsfonts, amscd, color}

\newtheorem{theorem}{Theorem}

\newtheorem{proposition}[theorem]{Proposition}

\newtheorem{corollary}[theorem]{Corollary}

\newtheorem{lemma}[theorem]{Lemma}

\newtheorem{remark}[theorem]{Remark}

\newtheorem{definition}[theorem]{Definition}

\newcommand{\R}{{\mathbb{R}}}
\newcommand{\Z}{{\mathbb{Z}}}
\renewcommand{\H}{{\mathbb{H}}}
\newcommand{\C}{{\mathbb{C}}}
\newcommand{\CP}{{\mathbb{C}}P}

\def\be{\begin{equation}}
\def\ee{\end{equation}}
\def\ba{\begin{array}}
\def\ea{\end{array}}
\def\vp{\varphi}

\begin{document}

\title{The Pontrjagin-Hopf invariants for Sobolev maps}
\author{Dave Auckly \thanks{{The first author was partially supported by NSF grants DMS-0204651 and DMS - 0604994.}}\and
Lev Kapitanski \thanks{{ The second author was partially supported by NSF grants DMS-0200670 and DMS-0436403.} }%
}                     

\date{
$$
\begin{array}{cc}
\hbox{Department of Mathematics} & \hbox{Department of Mathematics}\\
\hbox{Kansas State University} & \hbox{University of Miami}\\ \hbox{Manhattan, KS 66506, USA } & \hbox{Coral Gables, FL 33124, USA}
\end{array}
$$
}

\maketitle


\begin{abstract}
\end{abstract}


\section{Introduction}\label{intro}

Variational problems with topological constraints 
can exhibit many interesting mathematical properties. There are many open mathematical 
questions in this area. The diverse properties 
these problems possess give them the potential 
for different applications. We find 
variational problems with topological constraints 
in high energy physics, hydrodynamics, and material science, \cite{bmss,ArK,Taylor}.  Usually, the functional to be minimized represents energy 
and the topological constraint is that the map 
must be in a fixed homotopy class. 
An energy
functional dictates a natural class of maps -- the finite energy
maps. For some models 
finite energy maps include
discontinuous maps and this introduces new technical difficulties
into the homotopy theory. 
Many traditional arguments
have been designed for continuous or  smoother maps and do not work
for finite energy maps. Thus, one has to look for new approaches and
new interpretations that survive the lack of regularity. This may
lead to new geometric results and require more subtle analytic
techniques.
Even if all  
finite energy maps were continuous, one would 
want analytic expressions for 
complete homotopy invariants in order to apply 
the direct method in the calculus of variations. 
This is another challenge in geometry. 

This is a relatively new area of research with interesting interplay 
between geometry and analysis. The first steps have been made by studying the homotopy classes for Sobolev maps, see 
\cite{Burstall, White, Bethuel-1,Bethuel-2,BL,Brezis,HL}. 
One can view Sobolev maps as 
the maps with finite Sobolev energy aka the Sobolev norm. These studies are very important
because Sobolev spaces are a basic technical tool in analysis.

We were originally drawn into this area of research 
from our study of the Skyrme and Faddeev models. 
These are non-linear sigma models from high-energy 
physics \cite{Skyrme1, Skyrme2, Skyrme3, Faddeev, Faddeev2}. 
The Faddeev-Skyrme model has been generalized to encompass maps
between general Riemannian manifolds. The original Skyrme model was
for maps from $\R^3$ to the $3$-sphere, and the original Faddeev
model was for maps from $\R^3$ to the $2$-sphere.

In this paper
we concentrate on maps from $\R^3$ or a closed orientable
$3$-manifold to the $2$-sphere. (Maps into the $3$-sphere are easier
to analyze.) Continuous maps of this type are classified up to
homotopy by a primary invariant and a secondary invariant. In
\cite{AK2} we gave the first analytical description of the secondary
invariant for smooth maps. In this paper we extend the primary and secondary 
invariants to appropriate Sobolev maps and 
finite Faddeev energy maps and establish the
properties of these invariants expected from the smooth case. 
For continuous functions, the primary invariant can be viewed as an
obstruction to a certain lifting problem. We turn this around and use the
lifting problem to define the primary invariant for sufficiently
regular (but possibly discontinuous maps). 
In traditional homotopy theory, the homotopy class 
of a map is often studied by 
attempting to lift the map to higher levels 
of a Whitehead or a Postnikov tower. The lifting results in this paper can be viewed as a first step generalizing these techniques to Sobolev maps.  

Incidentally, lifting maps to maps taking values in a larger space has
the effect of replacing a variational problem with a new variational
problem with an additional differential equation constraint. The
hope is that it will be easier to analyze the new problem and
constraint than it is to analyze the old problem. This is indeed the
case with the Faddeev functional as we showed in \cite{AK2}.

We review the homotopy classification of 
$S^2$-valued maps 
in Section \ref{c0homotopy} and the definition of the Faddeev  and Skyrme functionals 
in Section \ref{fadsky} below. 
The local lifting result
appears in Section \ref{sec1}. It may be viewed as a non-linear
analog of the Poincar\'e lemma. This local result leads to the
global lifting criteria in Section \ref{sec2}. 
The local lifting result also leads to a 
definition of the primary invariant valid for finite energy maps. The secondary invariant compares two maps with the same primary invariant. Two such maps 
are related by an intertwining map. 
For the continuous case the degree of 
this intertwining map determines whether the 
two maps are homotopic. We extend this to finite 
energy maps by proving that the 
integral expression for the degree of such an intertwining map is an integer. 
We also show that changing the intertwining map 
changes this degree by an even multiple of the divisibility of the primary invariant as expected from the continuous case. 
Our results will generalize to other
spaces of maps with similar or better regularity. 

From what we prove
it follows, in particular, that for any finite energy map there is a smooth
map with the same invariants, and that two smooth maps have the
same invariants if and only if the maps are homotopic.

We reconsider minimization of the Faddeev functional  at the end of this paper.
We have made an effort to make this paper accessible to analysts,
topologists and mathematical physicists. The general outline of the
paper can be gleaned from the following table of contents:
\tableofcontents




\section{Review and examples}\label{rev}
This section contains a brief review of background material and several examples.


\subsection{Homotopy classification of $S^2$-valued
maps}\label{c0homotopy} The homotopy classification of maps $M\to
S^2$ has been known since the nineteen-thirties. First Hopf
\cite{Hopf} classified maps from $S^3$ into $S^2$ and defined a
complete integer-valued invariant that counts the linking number of
the inverse image of a pair of regular values. In modern notation
the result of Hopf is $\pi_3(S^2)=\mathbb Z$. Shortly thereafter
Pontrjagin \cite {Pontrjagin} studied the general case of maps from
a three-dimensional simplicial complex into $S^2$. When the
three-dimensional complex is an orientable manifold the result,
Pontrjagin writes, `may be formulated by means of the usual
homologies, which presents a certain advantage' (\cite[\S
4]{Pontrjagin}).  From this point forward we will restrict our
attention to closed orientable $3$-manifolds.

To set some useful notation we start by reviewing the easier case of
the homotopy classification of $S^3$-valued maps. A map $u:M\to S^3$
induces a map on the top cohomology $u^*:H^3(S^3;\Z)\to H^3(M;\Z)$.
Since the top cohomology of an oriented manifold is canonically
isomorphic to $\Z$ the map $u^*$ is just multiplication by an
integer. This integer is called the degree of the map and it is
denoted by $\hbox{deg}\, u$. The map $u$ is classified up to
homotopy by its degree. The intersection theory interpretation of
$\hbox{deg}\, u$ is the number of inverse images of a regular point
counted with sign. Using the deRham model one can describe the
degree of the map $u$ as the unique integer such that
$$
\int_Mu^*\alpha=(\hbox{deg}\, u)\int_{S^3}\alpha\,,
$$
for all top dimensional forms $\alpha$ on $S^3$. In particular one
has
$$
\hbox{deg}\, u=\int_Mu^*\omega_{S^3}\,,
$$
where $\omega$ is a normalized volume form on $S^3$, i.e.,
$\int_{S^3}\omega_{S^3}=1$. It is well-known but not obvious that
the three descriptions of the degree given above coincide
\cite{Bott-Tu}. The last description makes sense for possibly
discontinuous functions provided that the integral converges,
however it is far from obvious that the integral would still be an
integer (in fact, sometimes it is not). We will come back to this
point in great detail later.

Returning to the case of maps from $S^3$ to $S^2$ we see that the
Hopf invariant also has multiple descriptions. The description we
gave as the linking number of the preimages of two regular values is
the one arising from intersection theory. If a map $\varphi:\,S^3\to
S^2$ is sufficiently regular, there is a way to actually compute the
Hopf invariant of $\varphi$ by evaluating a certain integral. This
was found by J. H. C. Whitehead \cite{Whitehead} and goes as
follows. (See \cite{Bott-Tu} as well.) The pull-back of the
normalized volume form, $\varphi^*\omega_{S^2}$, is a closed
$2$-form on $S^3$, and, since $H^1(S^3)=0$, this form is exact,
i.e., $\varphi^*\omega_{S^2} = d\theta$, for some $1$-form $\theta$.
The result of Whitehead is
 \be\label{hopf}
\hbox{Hopf}\,(\varphi)\,=\,\int_{s^3}\theta\wedge d\theta\,. \ee
This is the expression analogous to the integral for the degree of
the map. In fact it is a de Rham description of a difference cocycle
arising from obstruction theory. In fact as we will explain in the
next subsection the Hopf invariant can be described as the degree of
a related map.

The final remark we should make about the Hopf invariant is that
since one can pick any normalized volume form $\omega_{S^2}$ and any
form $\theta$ satisfying $\varphi^*\omega_{S^2} = d\theta$, one can
pick nice forms. If one fixes a Riemanninan metric on $S^3$, then
among all $\theta$ satisfying $\varphi^*\omega_{S^2} = d\theta$
there is a unique one, $\theta^\varphi$, such that $\delta
\theta^\varphi = 0$, where $\delta$ is the adjoint of $d$ on
$1$-forms. This will be important when we generalize this invariant
to possibly discontinuous finite-energy maps.


\subsubsection{The classification}
Now return to the homotopy classification of maps from an arbitrary
$3$-manifold $M$ to $S^2$. Compared to the $\,S^3\to S^2\,$ case,
two new features arise. First, there is a new invariant given by the
induced map on second cohomology. Second, the Hopf invariant
generalizes into a secondary invariant. The secondary invariant is
an invariant of a pair of maps with the same primary invariant.  It
sometimes takes values in a finite cyclic group. The Hopf invariant
of a map is the secondary invariant of the pair consisting of the
map and a constant map.
\begin{theorem}[Pontrjagin]\label{one}
 Let $\,M\,$ be a closed, connected, oriented
three-manifold. To any continuous map $\,\vp\,$ from $\,M\,$
to $\,S^2\,$ one associates the pull-back
$\,\vp^*\mu_{S^2}\in H^2(M;\mathbb Z)$ of the orientation class
$\mu_{S^2}\in H^2(S^2;\mathbb Z)$.
Every cohomology class in $H^2(M;\mathbb Z)$
may be obtained from some map, and two maps with
different classes lie in different homotopy classes.
The homotopy classes of maps with a fixed class
$\,\alpha\in H^2(M;\mathbb Z)\,$
are in bijective correspondence with
$\,H^3(M; \mathbb Z)/(2\cdot\alpha\cup H^1(M;\mathbb Z))$.
\end{theorem}

For integral homology $3$-spheres (i.e. closed three-manifolds $M$
with $H_1(M)=0$), the homotopy classes of maps $M\to S^2$ are still
completely characterized by the Hopf number, which in the case of
sufficiently regular $\varphi$ can be computed via the same formula
(\ref{hopf}). If $H_1(M)\ne 0$, then the situation can be more
complicated. From Pontrjagin's description we see that the primary
invariant, $\,\alpha$,  is defined for individual maps as $\,\alpha
= \vp^*\mu_{S^2}$. If $\alpha = 0$, then
$$
\,H^3(M; \mathbb Z)/(2\cdot\alpha\cup H^1(M;\mathbb Z)) =
\,H^3(M; \mathbb Z) = \mathbb Z\,,
$$
and this is still the Hopf invariant. If $\alpha$ is not trivial and
is not pure torsion, then one must construct a secondary invariant.
The secondary invariant is a relative invariant, meaning that it
tells whether two maps with the same primary invariant are homotopic
or not. The homotopy invariants were originally discovered in the
framework of obstruction theory or intersection theory.

Using Poincar\'e duality we may identify the second cohomology with
the first homology to see the intersection theory interpretation.
Under this identification the primary invariant is just the homology
class of the inverse image of a regular point. The secondary
invariant is the relative framing of the inverse image under the
first map with respect to the inverse image under the second map.
More precisely, since we are assuming that the two maps have the
same primary invariant there is an oriented surface connecting the
inverse images of the two regular points. The relative framing
counts the number of intersections of this surface with the inverse
images of of two more regular points, one for each map. See
\cite{AS} for further exposition on this point. For integral
homology spheres the secondary invariant is just the Hopf invariant
which geometrically is the linking number of a pair of regular values.

In the framework of obstruction theory the primary invariant is the
obstruction to lifting the map $\vp:M\to S^2$ to a map $\Phi:M\to
S^3$ such that $\vp=\sigma\circ\Phi$, where $\sigma$ is the Hopf map
described later. The secondary invariant is just the class of the
difference cocycle. When the primary invariant vanishes so that
there is a lift, the secondary invariant reduces to the degree of the
lift $\Phi$.

It is instructive to consider a few examples. One could skip to the
examples in Section \ref{exmap} now. Before presenting the examples
we review the definition of the quaternions and several models of
$S^2$ that make it easier to construct examples. We also give a more
analytical description of the homotopy clasification. 


\subsubsection{Geometry of the quaternions}\label{quatrev}

Many important maps and forms related to the homotopy classification
of maps between a $3$-manifold and $S^2$ can be expressed in a
compact form using quaternionic notation. We denote a generic
quaternion by $q = q^0 + q^1\,{\bf i} + q^2\,{\bf j} + q^3\,{\bf k}$
and call $q^0$ the real part of $q$ and $q^1\,{\bf i} + q^2\,{\bf j}
+ q^3\,{\bf k}$ the imaginary part of $q$. Multiplication is
specified by requiring the quaternions to be a unital associative
algebra with
$$
{\bf i}^2={\bf j}^2={\bf k}^2={\bf ijk}=-1\,.
$$ 
Changing the sign of the imaginary part gives the
conjugate $\bar q$ of $q$, and one can check that
$\overline{pq}=\bar{q}\,\bar{p}$ on a basis. For {\it purely imaginary}  
quaternions, i.e., those with the real part zero, we have $\bar p = - p$. 

\noindent$\bullet$\ \ The set of all quaternions is denoted by $\H$.

\noindent The usual inner product on $\H$ is
$$
\langle p, q\rangle
= \frac12\,({\bar p}q + {\bar q} p)\,=\,p^0 q^0 + p^1 q^1 + 
p^2 q^2 + p^3 q^3.
$$
We denote the corresponding norm by $|p| = \langle p,\,
p\rangle^{1/2}$. 

\noindent$\bullet$\ \ The unit $3$-sphere
$S^3$ is identified with the  unit (norm $1$) quaternions. 

\noindent This is a Lie group  sometimes denoted 
$\textrm{Sp}(1)$. Its Lie algebra, 
${\mathfrak{sp}}(1)$, can be 
identified with the space of purely imaginary quaternions, $\R^3$,  
with the Lie bracket $[p,q] = pq - qp$. 
We will often use $x$ to
denote a purely imaginary quaternion. 
Viewing $x$ and $y$ as vectors in $\R^3$, we see that 
$\frac12\,[x, y]$ is  the usual cross-product
$x\times y$. Also, 
if, in addition, $x$ has norm one, 
then there is a useful decomposition of $y$ into a component parallel
to $x$ and a component perpendicular to $x$: 
\begin{equation}\label{decomp}
y=\langle y, x\rangle\, x \,+\,\frac12\,x\,[y,\,x]\,.
\end{equation}


\noindent$\bullet$\ \ The complex numbers embed into the quaternions as $a+b{\bf i}$.  The
set of quaternions $\H$ becomes a complex vector space with $\C$
acting on the left. This identifies $\C^2$ with $\H$ via
$(z,w)\mapsto z+w{\bf j}$. Since every complex vector space has a
canonical orientation this induces an orientation on $\H$. The
resulting complex orientation is given by $dq^0\wedge dq^1\wedge
dq^2\wedge dq^3$.


 The group of unit quaternions 
can be identified with
the group of special unitary matrices $\text{SU}(2)$ via the
identification of $\H$ with $\C^2$.  This association takes any unit
quaternion $q$ to the unitary map $p\mapsto p\bar q$. 
Here  we
take the conjugate of $q$ so that the $\text{SU}(2)$ action is a
left action. More concretely we have
\[ {\bf i} \mapsto
\begin{pmatrix}-i&0\\0&i \end{pmatrix},\qquad {\bf j} \mapsto
\begin{pmatrix}0&0\\-1&1 \end{pmatrix},\qquad {\bf k} \mapsto
\begin{pmatrix}0&-i\\-i&0 \end{pmatrix}.
\]
Other identifications are possible as well. 
The Lie algebra ${\mathfrak{sp}}(1)$ can be identified with the Lie algebra $\mathfrak{su}(2)$ via the
complex vector space structure on $\H$ used above. 

Some people prefer to think about SU$(2)$, others about $S^3$ and
others about Sp$(1)$. Of course all three are just different names
for the same thing. 
We choose to use quaternionic
notation because it is shorter.

\noindent$\bullet$\ \ We identify the
usual sphere $S^2$ with the unit sphere in the space of purely
imaginary quaternions. 

\noindent$\bullet$\ \ We identify $S^1$ with the unit quaternions
of the form $q^0 + q^1\,{\bf i}$. Thus $S^2\subset S^3$, $S^1\subset
S^3$, and $S^2\cap S^1 = {\bf i}\cup -{\bf i}$.


\noindent A quaternion-valued differential form is a combination
$$a=a^0+a^1{\bf i}+a^2{\bf j}+a^3{\bf k}\,,$$ and quaternionic
conjugation and multiplication (therefore the inner product and
commutator) extend to quaternion-valued differential forms. The
extension of conjugation is obvious and the multiplication of such forms
is given by:
$$
(a\wedge b)(X_1,\dots,X_k,X_{k+1},\dots
X_{k+\ell}):=\frac{1}{k!\ell!}\sum_\sigma (-1)^\sigma
a(X_{\sigma(1)},\dots,X_{\sigma(k)})b(X_{{\sigma(k+1)}},\dots
X_{{\sigma(k+\ell})})\,.
$$
If $\alpha$ and $\beta$ are real differential forms and $p$ and $q$
are quaternions then the wedge-product of the quaternion-valued forms
$p\,\alpha$ and $q\,\beta$ is given by
$(p\,\alpha)\wedge(q\,\beta)=p q\,\alpha\wedge\beta$ and extending this
linearly one recovers the definition of the quaternionic wedge-product of
forms.

\begin{remark}
Throughout the paper we will use $\omega_M$ to denote a normalized
volume form on a manifold $M$. Given a Riemannian metric on an
oriented manifold $M$ there is a unique compatible volume form
denoted by $d\,\hbox{vol}_M$  and specified by
$d\,\hbox{vol}_M(e_1,\dots,e_n)=1$ for any oriented orthonormal
basis $\{e_k\}$. A normalized volume form can then be constructed by
dividing by the volume of the manifold. Applying this procedure to
the low-dimensional unit spheres using the induced metrics and
orientations (outer normal first convention) gives the forms defined
below.
\end{remark}
\begin{definition}
The normalized volume forms on $S^1$, $S^2$ and $S^3$ are
\begin{small}
\begin{align*}
\,\omega_{S^1}\,&=\,\frac{1}{2\pi}\left( x^1\,dx^2
\,-\,x^2\, dx^1\right)\,\\
\,\omega_{S^2}\,&=\,\frac{1}{4\pi}\left( x^1\,dx^2\wedge
dx^3\,+\,x^2\,dx^3\wedge dx^1\,+x^3\,dx^1\wedge dx^2\right)\,\\
\,\omega_{S^3}\,&= \,\frac{1}{2\pi^2}\left( x^1\,dx^2\wedge
dx^3\wedge dx^4\,-\,x^2\,dx^1\wedge dx^3\wedge
dx^4\,+\,x^3\,dx^1\wedge dx^2\wedge dx^4\,-\,
x^4\,dx^1\wedge dx^2\wedge dx^3 \right)\,\\
\end{align*}
\end{small}
\end{definition}
In this last definition  $S^n$ is the unit sphere in $\R^{n+1}$ with
coordinates $(x^1,\dots,x^{n+1})$. Using the quaternionic
description of the low-dimensional spheres we have the following
lemma obtained from direct calculation.
\begin{lemma}\label{normform}
The normalized volume form on $S^1$ is given by
\be\label{volS1}
\omega_{S^1}=\frac{1}{2\pi{\bf i}}\,z^{-1}dz\,.
\ee
The normalized volume form on $S^2$ is given by
\be\label{volS2}
\omega_{S^2}=-\frac{1}{8\pi}\, x\, dx\wedge dx\,.
\ee
The normalized volume form on $S^3$ is given by
\be\label{volS3}
\omega_{S^3}=-\frac{1}{12\pi^2}\,\text{\rm Re}\,\left(q^{-1}dq\wedge
q^{-1}dq\wedge q^{-1}dq\right)\,.
\ee
\end{lemma}

\noindent{\bf Proof.}\ The volume form on $S^1$ is well known. For
$S^2$ begin by noting that $x\,dx\wedge dx$ is real. To see this
note that $|x|=1$ and $x+\bar{x}=0$ implies that $x^2=-1$, so
$x\,dx+dx\,x=0$ and
\[
\overline{x\,dx\wedge dx} = \overline{dx\wedge dx}\;\;\overline{x} =
- \overline{dx} \wedge \overline{dx} \,( - x) = dx\wedge dx\,x =
-dx\wedge x\,dx = x\,dx\wedge dx\,.
\]
Since
$x\,dx\wedge dx$ is real we can expand it keeping track of just the
real products to obtain the result.

For the $S^3$-case, notice that $|q|=1$ implies that $q^{-1}dq$ is
purely imaginary and $q^{-1}dq$ is invariant under left
multiplication by constants from $S^3$. It follows that we can just
do the comparison at $q=1$ where $q^{-1}dq=dq^1{\bf i}+dq^2{\bf
j}+dq^3{\bf k}$. \hfill $\square$

It is well known that the quotient $S^1\backslash\textrm{Sp}(1)$ is
homeomorphic to $S^2$.  In fact the homeomorphism is given by
$\widehat\sigma([q])={q}^{-1}{\bf i}{q}$. 
The double cover of the three-dimensional rotation group
$\textrm{SO}(3)$ may
be identified with $\textrm{Sp}(1)$. Given a unit quaternion $q$ define a
rotation $A_q:\R^3\to\R^3$ by $A_q(x)=qx{q}^{-1}$. Every rotation
arises from exactly two unit quaternions in this way. This elegant
description of the $3$-dimensional rotation group is one of the main
advantages of quaternionic notation. We see 
that the homeomorphism
$\widehat\sigma$ 
just sends the equivalence class $[q]$ to the image of ${\bf i}$
under the rotation associated to $q^{-1}$.

 Notice that the action of $S^1$ in the above quotient is by
constant multiplication. This is different from the $S^1$ subgroup
of $S^3$ that corresponds to the diagonal matrices in
$\textrm{SU}(2)$.  
The $2$-sphere is also
homeomorphic to one-dimensional complex projective space.  Recall
that one-dimensional complex projective space is
$\CP^1=\C^2-\{0\}/\sim$ where $(z,w)\sim (z\lambda, w\lambda)$.
Equivalence classes are denoted by $[z:w]$. Using the identification
of $\H$ with $\C^2$, the homeomorphism between
$S^1\backslash\text{Sp}(1)$  and $\CP^1$ is just the inclusion. Of
course $\CP^1$ is also homeomorphic to the extended complex numbers
$\overline\C$. Here the identification takes $z$ to $[z:1]$ and
$\infty$ to $[1:0]$. The map from $S^3$ to $S^2$ given by
composition of projection with the homeomorphism is an important map
known as the Hopf map.
\begin{definition}
The Hopf map or Hopf fibration $\sigma:S^3\to S^2$ is given by
$\sigma(q)={q}^{-1}{\bf i}{ q}$.
\end{definition}
\begin{remark}
The Hopf map can be described in several different ways. A second
way to describe the same map is as the map from $S^3$ viewed as the
unit sphere in $\mathbb{C}^2$ to $\CP^1$ taking $(z,w)$ to 
$[z:w]$
or, using the extended complex plane, to $z/w$.
\end{remark}

The next subsection contains a series of computations leading to the
Hopf invariant of the Hopf map. Many of the intermediate formulas 
will be used repeatedly throughout the paper.


\subsubsection{Hopf invariant of the Hopf map}\label{hopfhopf}

The definition of the Hopf invariant does not depend on the choice
of normalized volume form. (One can see this by an elementary
application of de Rham theory.) We choose the standard volume form.

\begin{lemma}\label{dtheta}
Consider the Hopf map $\sigma :\,S^3\to S^2$ given by 
$q\mapsto x = q^{-1}{\bf i}q$. Define 
\[
a = q^{-1}dq
\]
and 
\begin{equation}\label{theta-def}
\theta = {1\over 2\pi}\,\langle a, x\rangle = {1\over 2\pi}\,\langle q^{-1}dq,q^{-1}{\bf i}q\rangle\,.
\end{equation}
Then  
\begin{equation}\label{theta-S2}
\sigma^*\omega_{S^2}=d\theta\,
\end{equation} 
and  
\begin{equation}\label{theta-S3}
\omega_{S^3} = -\frac{1}{12\pi^2}\,\text{\rm Re}\,\left(a^{\wedge 3}\right) = \theta\wedge\sigma^*\omega_{S^2} = \theta\wedge d\theta\,. 
\end{equation}
\end{lemma}


\noindent{\bf Proof.}\ \ When 
 $x = \sigma(q)=q^{-1}{\bf i}\,q$, we have 
\begin{equation}\label{bradif}
dx=-q^{-1}dq\,x+x\,q^{-1}dq=[x,a]\,.
\end{equation}
Then 
\begin{align}\label{brasq}
- 8 \pi\, \omega_{S^2} = x\,dx\wedge dx &=x[x,a]^{\wedge 2}\notag\\
&=-axa+a^{\wedge 2}x+xa^{\wedge 2}+xaxax = 
4\,\textrm{Re}\left(a^{\wedge 2}x\right)\,,
\end{align}
where in the last line we used the fact that $x\,dx\wedge dx$ is real to
cycle and combine terms. On the other hand, 
\begin{align}\label{dpar}
2 \pi\,d \langle a, x\rangle = d\langle q^{-1}dq,q^{-1}{\bf i}q\rangle&=d\langle dqq^{-1},{\bf
i}\rangle=\langle dqq^{-1}\wedge dqq^{-1},{\bf i}\rangle\notag\\
&=\langle a^{\wedge 2},x\rangle = - \textrm{Re}\left(a^{\wedge 2}x\right) = - \frac14\,x\,dx\wedge dx\,.
\end{align}
This proves that $d\theta=\sigma^*\omega_{S^2}$. 
To prove (\ref{theta-S3}) we work backwards with the last
equation from Lemma \ref{normform}, i.e.,
\[
\omega_{S^3} = - \frac{1}{12\pi^2}\,\hbox{Re}\,\left(a^{\wedge 3}\right)\,.
\]
Splitting $a$ into components
parallel and perpendicular to $x$ using equation (\ref{decomp}), 
we compute:
\[
a^{\wedge 3} = \left(\langle a,x\rangle
x + \frac12\,x [a,x]\right)^{\wedge 3} = 
\frac34\,\langle a,x\rangle\,x\wedge [a, x]^{\wedge 2} + \frac18\,
\left(x\,[a, x]\right)^{\wedge 3}\,.
\]
Here we used the fact that the form $\langle a,x\rangle$ is real 
and that $[a, x] x = - x [a,x]$ and $x^2 = -1$. 
A useful observation is that
\[
\hbox{Re}\,\left(x\,[a, x]\right)^{\wedge 3}\,=\,0\,, 
\]
because we
can cycle the terms in the real part without picking up signs, and because $x^2 = -1$. 
It follows that 
\begin{equation}\label{a^3}
\omega_{S^3} = 
-\frac{3}{4\cdot12\pi^2}\,\hbox{Re}\,\left(\langle a,x\rangle
x\wedge [a,x]^{\wedge 2}\right)
= -\frac{1}{8\pi}\,x[a,x]^{\wedge 2}\wedge\frac{1}{2\pi}\langle a,x\rangle
=\theta\wedge\sigma^*\omega_{S^2}\,.
\end{equation}
End of proof. \hfill$\square$%

\bigskip


\begin{corollary}\ \ 
$\hbox{\rm Hopf}\;(\sigma)\,=\,1$. 
\end{corollary}


\bigskip

There are a number of formulas that we will use in computations
throughout the rest of of this paper. All of them can be verified by
direct computation. We have collected them in the following lemma.

\begin{lemma}\label{formulas}
The following formulas are true when $|q|=1$ and 
$x,y\in{\mathfrak{sp} (1)}\cong\R^3$.
\begin{equation}\label{adinvariance}
\langle q^{-1}xq,q^{-1}yq\rangle=\langle x,y\rangle,
\end{equation}
The following formulas are true when $x=q^{-1}{\bf i}q$ and
$a=q^{-1}dq$.
\begin{equation}\label{dpar}
d\langle q^{-1}dq,q^{-1}{\bf i}q\rangle=- \frac14\,x\,dx\wedge dx,
\end{equation}
\begin{equation}\label{a^3}
\langle a, a^{\wedge 2}\rangle=-\text{\rm Re}\,(a^{\wedge
3})=-\frac34\;\text{\rm Re}\,(\langle a,x\rangle\; x\,[a,x]^{\wedge 2}).
\end{equation}
\end{lemma}


\subsubsection{Analytic description of homotopy for
maps $M^3\to S^2$} 

Using obstruction theory or intersection theory
is not a suitable method to keep track of the homotopy class of a
map for the functions with relatively little regularity that are
encountered in the variational problems that we considered. As far
as we know, until our recent paper \cite{AK2}, there were no
analytically suitable tools to distinguish all homotopy classes of
maps from $M$ to $S^2$ even for simple examples such as $S^2\times
S^1$, $T^3$ and $S^3/\Z_2$. Developing suitable tools was one of the
goals of our previous paper \cite{AK2}. That paper also contains a
new proof of Pontrjagin's theorem. The tools developed in \cite{AK2}
were sufficient to establish the existence of minimizers of the
Faddeev model in each sector, but a couple of important questions
were left unanswered. We address those questions in this paper. Here
is a summary of our description from \cite{AK2}.

\begin{theorem}\label{our}
Given a (smooth) map $\varphi:\,M\to S^2$, the primary homotopy
invariant is the class $\varphi^*\mu_{S^2}\in H^2(M;\mathbb Z)$. The
map $\,\eta \mapsto (\varphi^*\mu_{S^2}\cup \eta)[M]\,$ from
$\,H^1(M;\mathbb Z)\,$ to $\,\mathbb Z\,$ is a group homomorphism,
and therefore  has image $\,m\,\mathbb Z\,$ for some $\,m\,$
determined by the class $\,\varphi^*\mu_{S^2}$. All maps $\psi:
\,M\to S^2$ with the same second cohomology class $\,\psi^*\mu_{S^2}
= \varphi^*\mu_{S^2}\,$ are obtained in the form
$$
\psi(x)\,=\,\Phi(x)\,\varphi(x)\,\Phi(x)^{-1}\,,
$$
where $\Phi$ is a map from $M$ into $S^3$. Furthermore, two maps
$\,\Phi_1\,\varphi\,\Phi_1^{-1}\,$ and
$\,\Phi_2\,\varphi\,\Phi_2^{-1}\,$ are homotopic if and only if
$\,\hbox{deg}\;\Phi_1\,\equiv \hbox{deg}\;\Phi_2\;
(\,\hbox{\rm mod}\; 2
m)$.
\end{theorem}

For the special case of maps $S^3\to S^2$ the primary homotopy
invariant is trivial, so this theorem implies that every map
$\varphi:S^3\to S^2$ is related to the constant map ${\bf i}:S^3\to
S^2$ via a lift $\Phi:S^3\to\hbox{Sp}(1)$ such that
$\varphi(x)\,=\,\Phi^{-1}(x){\bf i}\Phi(x)=\sigma\circ\Phi(x)$. This
is exactly what one would expect from the interpretation of the
primary invariant as a lifting obstruction. Notice that it is not
the case that every map from a general $3$-manifold to $S^2$ factors
through the Hopf map in this way (the primary invariant is the
obstruction to the existence of such a lift). This is why we had to
introduce maps intertwining a pair of maps with the same primary
invariant in our analytic description. The only homotopy invariant
in the special case of maps $S^3\to S^2$ is therefore the degree of
the map $\Phi$. As we reviewed at the start of this section such
maps are classified up to homotopy by the Hopf invariant. Thus one
would expect that Hopf$(\varphi) = - \hbox{deg}\, \Phi$. This is
indeed the case and the proof is given in Theorem \ref{ourhopf}
below.

The importance of the representation 
$\varphi(x)\,=\,\Phi^{-1}(x){\bf i}\Phi(x)$ 
has been recognized since the
work of Hurewicz \cite{Hur}. That a continuous map $\vp:\,S^3\to
S^2$ can be written as $\Phi^{-1}\,{\bf i}\,\Phi$ was proved, for
example, in Pontrjagin's paper \cite[Lemma 3]{Pontrjagin}. We gave a
different proof for the smooth case in our earlier paper on the
existence of minimizers of the Faddeev functional \cite{AK2}. R.
Hardt and T. Rivi\`ere gave an analytic proof of the existence
of the lift $u$ in the case of a $C^\infty$ map $\vp$ (\cite[Lemma
2.1]{HR}; their notation is different from ours). They also noted
that since $W^{1,3}(S^3, S^2)$ maps can be approximated by
$C^\infty$ maps, the Hopf number of such a map is well-defined
either by approximation or as the degree of the corresponding
$W^{1,3}$ lift, the latter approach has been used by T. 
Rivi\`ere in \cite{R1}.

In their paper \cite{LY}, dealing with the Faddeev energy
functional, Fanghua Lin and Yisong Yang also invoke a `Hopf lift'
to address  integrality of the Hopf number (note that they
denote by $u$ the map into $S^2$ and by $\bar{u}$ its lift). The
idea behind the integrality proof is quite simple and goes as
follows. Assuming the existence of a lift $\Phi$ such that
$\vp=\Phi^{-1}\,{\bf i}\,\Phi$ one has $\vp=\sigma\circ\Phi$, where
$\sigma$ is the Hopf map. This implies that
$\vp^*\omega_{S^2}=\Phi^*\sigma^*\omega_{S^2}=\Phi^*d\theta=d\Phi^*\theta$,
where $\theta=\frac{1}{2\pi}\langle q^{-1}dq,q^{-1}{\bf i}q\rangle$.
Thus we can take $\Phi^*\theta$ as the form in the definition of the
Hopf invariant to compute
\[
\text{Hopf}(\vp) =  \int_{S^3}\Phi^*\theta\wedge
d\Phi^*\theta = \int_{S^3}\Phi^*(\theta\wedge d\theta)=-\int_{S^3}\Phi^*\omega_{S^3}\\
 = \,\text{deg}\;\Phi\,.
\]
This explanation extends easily to the maps $\vp\in W^{1,3}$ since
such maps can be approximated in the $W^{1,3}$ norm by smooth maps.

Adapting this argument for maps $\vp$ with finite Faddeev energy is
more difficult. The problem is with the assertion that the degree of
the lift is integral. F. Lin and Y. Yang consider maps from
$\mathbb{R}^3$ into $S^2$ and $S^3$. They refer to the paper
\cite{EM} of M. Esteban and S. M\"uller for the integrality
of the degree of the lift. In \cite{EM} the integral
$\int_{\mathbb R^3}u^*\omega_{S^3}$ is proved to be an integer for the maps
$u:\,\mathbb R^3\to S^3$ with {\it finite Skyrme energy}, i.e.,
$u^{-1}du\in L^2$ and $u^{-1}du\wedge u^{-1}du\in L^2$. There is no
justification in the paper of F. Lin and Y. Yang \cite{LY}
that the lift $\bar{u}$ they construct does have finite Skyrme
energy. 
Indeed,
Example 8 from Section \ref{sec1} 
shows that lifts of finite Faddeev
energy maps need not have finite Skyrme energy. 
The lift $\bar{u}$ in \cite{LY} satisfies $\bar{u}\in\dot W^{1,2}(\mathbb R^3,
S^3)$ and $\bar{u}^*\omega_{S^3}\in L^1(\mathbb R^3)$, but as one
can see from the function from Example 6 in Subsection \ref{sample}
this is not enough to justify integrality.

In Proposition \ref{rep}, Section \ref{sec1} we give an alternative
construction of a lift that does have  nice analytic
properties, and 
we  prove that $\hbox{Hopf}\,(\vp)\,=\,\hbox{deg}\,\Phi$
is an integer. We use these properties later when we consider the
more complicated situation of maps from an arbitrary closed
three-dimensional manifold $M$ into $S^2$, define the primary
homotopy invariant, and compare pairs of such maps with the same
primary invariant.

Our description of the homotopy classification together with the
formula (Lemma \ref{normform})  for the normalized volume form
prompts the following definition of a numerical secondary homotopy
invariant \cite{AKS}. Given $\vp, \psi:M\to S^2$ such that
$\vp^*\mu_{S^2}=\psi^*\mu_{S^2}$ define
\begin{equation}\label{upsilon}
\Upsilon(\vp,\psi) = 
- \frac{1}{12\pi^2}\int_M\hbox{Re}\,\left(
\Phi^{-1}d\Phi \wedge \Phi^{-1}d\Phi \wedge \Phi^{-1}d\Phi
\right)\;\qquad\hbox{mod}\;(\,2m_\psi),
\end{equation}
where $\Phi$ is the map intertwining $\varphi$ and $\psi$, i.e.,
$\,\psi\,=\,\Phi\,\varphi\,\Phi^{-1}$,  and $m_\psi$ is the
divisibility of the class $\psi^*\mu_{S^2}$.  The following lemma is
used in the result first stated in \cite{AKS} that relates the
Hopf invariant and $\Upsilon$ in case the class $\psi^*\mu_{S^2}$ is
torsion.

\begin{theorem}\label{ourhopf}
Let $N$ be the smallest positive integer such that
$N\,\varphi^*\mu_{S^2}=0$. We have
$$
\hbox{\rm Hopf}\,(\varphi) = \frac{1}{N^2}\; \Upsilon(\varphi_N,{\bf
i}),
$$
where $\varphi_N$ is the composition of $\varphi$ with the map
$z\mapsto z^N$ of $S^2=\C\cup\{\infty\}$, and ${\bf i}$ is the
constant map from $M$ to ${\bf i}\in S^2$.
\end{theorem}

\smallskip

\noindent{\bf Proof.} Define a map $p_N:{\mathbb C}P^1\to {\mathbb
C}P^1$ by $p_N([z:w])=[z^N,w^N]$. Since ${\mathbb C}P^1$ is
diffeomorphic to $S^2$ we may consider $p_N$ as a smooth map of
degree $N$ on $S^2$. This implies that there is a smooth $1$-form
$\alpha$ on $S^2$ such that
$\frac{1}{N}p_N^*\omega_{S^2}=\omega_{S^2}+d\alpha$.

Given $\varphi$ as in the proposition, the map
$\varphi_N=p_N\circ\varphi$ has the same regularity as $\varphi$ and
satisfies $\varphi_N^*\mu_{S^2}=0$. By Theorem \ref{our} there is a
map $\Phi:M\to S^3$ satisfying $\varphi_N=\Phi^{-1}{\bf i}\Phi$. 
Denote 
\[
\theta_N = \frac{1}{2\pi N} \langle \Phi^{-1}d\Phi,\varphi_N\rangle\,.
\]
Equation (\ref{dpar}) in Lemma \ref{formulas} together with the
quaternionic form of $\omega_{S^2}$ (Lemma \ref{normform})  shows that 
$ d\theta_N = \frac{1}{N}\varphi^*_N\omega_{S^2}$, 
which is equal to 
$\frac{1}{N}\vp^*p_N^*\omega_{S^2} 
= \varphi^*\omega_{S^2}+d\varphi^*\alpha\,$. Thus 
\[
d \theta_N = \varphi^*\omega_{S^2}+d\varphi^*\alpha
\]
and 
\[
\vp^*\omega_{S^2}=d\theta\,,\quad\text{where}\quad 
\theta = \theta_N - \vp^*\alpha\,.
\]
Now, 
\[
\begin{aligned}
\theta\wedge d\theta & = \left(\theta_N - \vp^*\alpha\right) 
\wedge d\,\left(\theta_N - \vp^*\alpha\right) \\  
 & = 
\theta_N\wedge d\theta_N + d\left(\theta_N\wedge \vp^*\alpha\right)
 + \vp^*\alpha\wedge d\vp^*\alpha - 2 \vp^*\alpha\wedge d\theta_N .
\end{aligned}
\]
The terms $\vp^*\alpha\wedge d\vp^*\alpha
- 2 \vp^*\alpha\wedge d\theta_N $ 
vanish, because they are the pull-backs of $3$-forms on $S^2$. 
Applying Lemma 
\ref{dtheta}, formula (\ref{theta-S3}), we obtain  
\[
\theta_N\wedge d\theta_N = - \frac{1}{N^2}\frac{1}{12\pi^2}\,
\hbox{Re}((\Phi^{-1}d\Phi)^3)\,.
\] 
Finally,   
\[
\hbox{Hopf}\,(\varphi) =  \int_M \theta\wedge d\theta = 
-\,\frac{1}{N^2}\frac{1}{12\pi^2}\int_M\hbox{Re}\,((\Phi^{-1}d\Phi)^3) =
\frac{1}{N^2}\Upsilon(\varphi_N,{\bf i}).
\]
\hfill$\square$%

\begin{figure}
\hskip115bp \epsfig{file=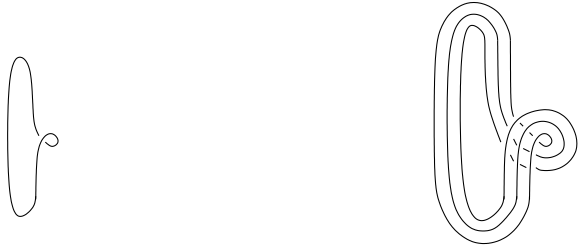, width=3truein}
 \caption{Change in Hopf invariant from composition with a degree
$3$-map}\label{n2fig}
\end{figure}

The factor of $N^2$ in this last theorem can easily be understood
from the linking number description of the Hopf invariant. Figure
\ref{n2fig} displays the inverse image of a regular point under the
Hopf map on the left (the inverse image of a second regular point
would just be a parallel copy of the curve). The linking number
between these two inverse images can be seen as the number of
crossings in the figure. The inverse image of one regular point
under the composition of a degree $3$ self-map of $S^2$ with the
Hopf invariant is drawn on the right. Clearly the linking number
multiplies by $3^2$.

The open questions from \cite{AK2} that we address in this paper are
related to generalizing the description of the homotopy
classification given in Theorem \ref{our} to finite energy maps.
Since finite energy maps may be discontinuous, the correct
definition of $\varphi^*\mu_{S^2}$ is not clear. {\it A priori}, the
integral $\Upsilon$ only takes integer values for smooth maps, so
one would like to know that it  takes integer values for
finite-energy maps. We resolve these questions by defining
$\varphi^*\mu_{S^2}$ for finite energy maps in Section \ref{sec2}
and proving that $\Upsilon$ only takes integer values in Section
\ref{intsec}. We next present a number of examples of maps.


\subsubsection{Examples}\label{exmap}

Maps from $3$-manifolds to $S^3$ are classified by the degree, but
it is instructive to construct specific representatives. 
For a geometer, the
easiest way to construct a degree one map from an arbitrary
$3$-manifold $M$ to $S^3$ is to take a cell decomposition of $M$ with only one $3$-cell and collapse the $2$-skeleton
$M^{(2)}$. 
Clearly, the inverse image of a generic
point under this map consists of just one point. One can change the
orientation to make the degree either $+1$ or $-1$. Maps of any
degree can then be obtained by postcomposition with self maps of
$S^3$ of the right degree. The maps constructed in this way will be
continuous but are unlikely to be smooth. Of course a well-known
theorem of Whitehead tells us that we can approximate any continuous
map arbitrarily closely by a smooth map in the same homotopy class.

Here is an alternate construction that produces smooth maps to
begin with. The first step is to produce a map $\R^3\to S^3$ 
which can be patched into any $3$-manifold later. The 
map $\R^3\to S^3$ is a modification of the stereographic projection 
taking the exterior of the unit ball to a pole. It is a composition 
of inversion ($x \mapsto |x|^{-2} x$), rescaling ($x\mapsto \rho(|x|)x$) 
via a cut-off 
function $\rho$ equal to zero
on $[0,1]$ and equal to one on $[2,\infty)$, and stereographic projection ($x \mapsto (1 - |x|^2 + 2x)(1 + |x|^2)^{-1}$), i.e.,
\[
x \mapsto 
\frac{1 - |\rho(|x|^{-1})|x|^{-2} x|^2 + 2 \rho(|x|^{-1})|x|^{-2} x}
{1 + |\rho(|x|^{-1})|x|^{-2} x|^2}\,.
\]  
Given any local chart in a $3$-manifold $M$, i.e., 
a  copy of $\R^3$ embedded in  $M$, 
we can 
smoothly map $M$ to $S^3$ by using this map on the copy of $\R^3$
and mapping the rest to the pole. This will be a smooth map of
degree $+1$ or $-1$ depending on the orientation of the copy of
$\R^3$. Of course this is just a smoothing of the collapsing map
first described.

Of course the same construction will give a degree
one map from any closed $n$-manifold to $S^n$.

Given a finite collection of points in $M$ one can take disjoint
copies of $\R^3$ around each point and map each to $S^3$ with our 
standard map and map the rest of $M$ to the pole to obtain a map of
arbitrary degree. These maps can be viewed as collections 
of lumps representing particles and anti-particles. Moving 
the lumps or canceling/adding lump-antilump pairs does not change the homotopy class of the map. In this interpretation the degree 
is the net number of particles. 

\bigskip

We now turn to the case of $S^2$-valued maps. By the end of our
examples we will have a picture of $S^2$-valued maps similar to our
picture of $S^3$-valued maps. The Hopf map itself is the most
important example of a map from a $3$-manifold to $S^2$. The Hopf
map is just the result of conjugating the constant map by a degree
one map from $S^3$ to $S^3$. According to our general theory, one can obtain 
representatives of every homotopy class of maps by
conjugating one map with each primary invariant by maps from $M$ to
$S^3$.

\bigskip

\noindent{\bf Example 0.} \ \ 
$M=S^3$. One can check that the map
$q\mapsto q^n$ has degree $n$. It follows that every map from $S^3$
to $S^2$ is homotopic to exactly one of the form $q\mapsto
q^{-n}{\bf i}q^n$.

\bigskip

\noindent{\bf Example 1.} $M = S^2\times S^1$. The second cohomology
group of $S^2\times S^1$ is the infinite cyclic group generated by
$p^*\mu_{S^2}$ where $p:\,S^2\times S^1\to S^2$ is projection and
$\mu_{S^2}$ is the generator of $H^2(S^2;\Z)$. A map with primary
invariant $m\,p^*\mu_{S^2}$ is constructed by composition of
projection with a degree $m$ self map of $S^2$. Let $p_m$ be such a
map. Let $q:\,S^2\times S^1\to S^3$ be a degree one map. Every map
$S^2\times S^1\to S^3$ is homotopic to exactly one of
\[
q^{-n}\,p_m\,q^n\quad \text{for}\quad n=1,\cdots, 2\,|m|\ne 0
\quad\text{or}\quad q^{-n}\,p_0\,q^n.
\]

\medskip

\noindent{\bf Example 2.} $M = T^3$, the $3$-torus.  We follow the same
outline to construct maps here that we did in the $S^2\times S^1$
case: we first construct maps with given primary invariant, and then
twist these to obtain the rest.

Represent the $3$-torus as ${\mathbb R}^3/{\mathbb Z}^3$. The second
cohomology of $T^3$ with integer coefficients may then be viewed as
the following subset of the second deRham cohomology,
$\{(m_1,m_2,m_3) = m_1\,dx^2\wedge dx^3+m_2\,dx^3\wedge
dx^1+m_3\,dx^1\wedge dx^2\}$ with $m_k$ integers. Maps with given
primary invariant can be constructed as the composition of a
`linear' projection from $T^3$ to $T^2$ followed by the collapse of
the $1$-skeleton taking $T^2$ to $S^2$. Viewing $T^2$ as $\R^2/\Z^2$
an integer $2\times 3$ matrix gives a well defined map from $T^3$ to
$T^2$. The induced map on the second cohomology is just given by the
determinants of the $2\times 2$ minors of the transpose of this
matrix. It is an elementary exercise in number theory to see that
any triple of integers arises as the determinants of the minors of
such a matrix. For example,
\[
\left[\begin{array}{rrr}
3 & 1 & 2 \\
-5 & 0 & -2
\end{array}\right]\,,
\]
corresponds to the primary invariant $(-2,-4,5)$. Let
$p_{(m_1,m_2,m_3)}$ denote the map with primary invariant
$(m_1,m_2,m_3)$. Let $q:T^3\to S^3$ be a degree one map, then every
map $T^3\to S^2$ is homotopic to exactly one of
\[
q^{-n}p_{(m_1,m_2,m_3)}q^n\quad \text{for}\quad
n=1,\cdots,2\;\text{gcd}{(m_1,m_2,m_3)}\ne 0\quad\text{or}\quad
q^{-n}p_{(0,0,0)}q^n\,.
\]

\medskip

\noindent{\bf Example 3.} $M = S^3/\Z_2$, the $3$-dimensional
projective space.  The interesting thing about this case is that the
primary invariant takes values in $H^2(S^3/\Z_2)\cong\Z_2$. Every
possible primary invariant is taken by a map  of the form
$\varphi_k: S^3/\Z_2\to S^2$, given by $\varphi_k([q]):=q^k{\bf
i}q^{-k}$. The primary invariant is trivial if and only if $k$ is
even and equal to the non-zero element of $H^2(S^3/\Z_2)\cong\Z_2$
when $k$ is odd. We can see this because $\vp_0$ is the constant map
and each $\vp_{2\ell}$ is related to $\vp_0$ by conjugation by a map
from $S^3/\Z^2\to S^3$. Likewise all $\vp_{2\ell+1}$ have the same
primary invariant and one sees that the primary invariant of $\vp_1$
is represented by the inverse image of a regular point which is
exactly the geometric image of a great circle in $S^3$ in
$S^3/\Z_2$. This represents the nonzero element of $H_1(S^3/\Z_2)$.

To continue and obtain all homotopy classes of maps we just have to
intertwine (conjugate) one of each primary invariant with maps from
$S^3/\Z_2$ to $S^3$ having all possible degree. In this case one
should notice that the map $[q]\mapsto q^2$ is well defined and has
degree one since the inverse image of one point is plus and minus
some point which represents just one point in $S^3/Z_2$. Thus
$\vp_k$ represent all homotopy classes of maps.

\medskip 

There is a uniform way to describe all homotopy classes of maps from
any $3$-manifold to $S^2$. Start with an embedded copy of $S^1\times
\R^2$ in the $3$-manifold. Project any point in this solid torus to
$\R^2$ then map $\R^2$ to $S^2$ by the map analogous to the map that
we constructed from $\R^3\to S^3$ at the start of the section. Map
the rest of the $3$-manifold to the pole. This is the
Pontrjagin-Thom construction. In general any homotopy class of maps
from a manifold to a sphere can be represented in this way. The
primary invariant of a $S^2$-valued map is the Poincar\'e dual of
the core ($S^1\times\{0\}$) of the solid torus. Roughly, the value
of this $2$-cohomology class on a $2$-cycle is the number of points
in the intersection of the cycle with the core of the solid torus.
The secondary invariant measures the relative number of twists
between two solid tori representing the same primary invariant. If
an energy density is zero when the derivative is zero, the core of
such a solid torus will be a vortex line (the inverse image of a regular point, a codimension $2$
submanifold around which energy concentrates). 
Of course multiple
vortex lines can be inserted into any $3$-manifold. 
If the map
evolves in a way preserving the homotopy class, the vortex lines
will persist. More precisely the homology class generated by the
vortex lines will be constant. There is some numerical evidence that minimizers of the Faddeev 
functional concentrate along such vortex lines, 
\cite{Faddeev-Niemi}.

Returning to the examples from this section we can see the vortex
line structure by taking the inverse image of a regular point. For
Example 1 ($S^2\times S^1$) the inverse image of a regular point
under a degree $m$ self map of $S^2$ is $m$ points which then give
$m$ copies of the fiber $\{\text{point}\}\times S^1$ as vortex
lines. As conjugation by a quaternion corresponds to rotation of
$\R^3$, conjugating these maps just `twists' the vortices. Similarly,
for Example 2 ($T^3$), the inverse image of a regular point just
produces $\text{gcd}(m_1,m_2,m_3)$ parallel copies of a circle in
the $(m_1,m_2,m_3)$-direction.


\subsection{Sobolev maps between manifolds}

Over the last thirty years Sobolev maps between manifolds have
become an indispensable part of the mathematical arsenal of
topologists, geometers, analysts, and theoretical physicists. Many
facts about Sobolev maps can be carried over from the well developed
theory of Sobolev spaces on $\mathbb R^n$, \cite{Adams}. However,
there are interesting analytical problems specific to maps between
manifolds that attract considerable attention, see \cite{Brezis} and
references therein. Not surprisingly the analytical complications
arise when there are topological obstructions that are not respected
by the spaces in question.

Recall, that for a domain $\Omega$, $\,s\,$ a non-negative integer
and $\,p\ge 1$, a function $\,f\,$ belongs to $\,W^{s,p}(\Omega)\,$
if it is Lebesgue measurable on $\,\Omega\,$ and $\,f\,$ itself and
all of its partial derivatives of order up to $\,s\,$ are integrable
with power $\,p$ over $\,\Omega$. The partial derivatives can be
understood in terms of distributions. Identifying the functions that
differ on locally negligible (measure $0$) sets, and introducing the
norm $\, \|f\|_{W^{s,p}}\,=\,\sum_{|\alpha|\le s} \|\partial^\alpha
f\|_{L^p}(\Omega)$, makes $\,W^{s,p}(\Omega)\,$ into a Banach space.
This definition extends easily to vector-valued functions: an
$\,\mathbb R^N$-valued function $\,f\,$ belongs to
$\,W^{s,p}(\Omega; \mathbb R^N)\,$ if each of its $\,N\,$ components
$\,f^1\,$ through $\,f^N\,$ belongs to $\,W^{s,p}(\Omega)$. The norm
is the sum of the norms of the components. Also recall that it is
traditional in differential topology to define smooth functions on
closed subsets as those that are restrictions of smooth functions
from open subsets containing the given closed subset.

Another extension is to functions on a manifold. 
Let $\,X\,$ be a smooth, closed, connected manifold of dimension $n$. 
Let $\{(U_i, \chi_i)\}$ be a finite atlas, and let 
$\{\vartheta_i\}$ be a partition of unity corresponding to the cover 
$\{U_i\}$. The Sobolev space  $\,W^{s,p}(X, \R^N)$ is comprised 
of functions $f:\,X\to \R^N$ such that each product  
$f_i = \vartheta_i\cdot f$ is in $\,W^{s,p}(X,\,\R^N)$, i.e., the composition 
$f_i\circ \chi^{-1}$ is in $\,W^{s,p}(\R^n,\,\R^N)$. We set 
\[
\|f\|_{W^{s,p}(X,\,\R^N)} = \sum_i \|f_i\circ \chi_i^{-1}\|_{W^{s,p}(\R^n,\,\R^N)}\,.
\]
When the cover $\,\{U_i\}\,$ and/or partition of unity
$\,\{\vartheta_i\}\,$ are changed, the Sobolev spaces do not change, but the norms change to the equivalent ones. The space $\,W^{0,p}(X,;\mathbb R^N)\,$
is identified with the Lebesgue space 
$\,L^p(X,;\mathbb R^N)$.

There is a more traditional definition of Sobolev spaces on a
manifold. Choose a smooth Riemannian metric on $\,X$. This, in
particular, defines a measure, $\,d\hbox{vol}$, on $\,X$. The
$\,L^p\,$ spaces are defined with respect to this measure. The
Sobolev space $\,W^{s,p}$, when $\,s\,$ is a positive integer, can
be defined using covariant derivatives as the closure of smooth
functions in the norm $\,\sum_{\ell=0}^{\ell=s} \|\nabla^\ell
f\|_{L^p}$. For non-integral $\,s$, the spaces $\,W^{s,p}\,$ are
defined, usually, via real interpolation between spaces $\,L^p\,$
and $\,W^{m,p}$, and are, in fact, Besov spaces, $\,B^{s;\,p,p}$,
see \cite{Adams}. For  Sobolev spaces on Riemannian manifolds, see
\cite{Aubin,Hebey}.

Since we will be discussing and generalizing results from
differential topology to Sobolev-valued differential forms, it is
worth spelling out the relevant definitions here. The covariant
derivative also acts on differential forms, so Sobolev spaces of
forms $ W^{s,p}(\wedge^kX) $ are just the completion of the space of
smooth forms $C^\infty(\wedge^kX)=\Omega^k(X)$ with respect to the
norm
$$
\|\alpha\|^p_{W^{s,p}}:=\sum_{m=0}^\infty\int_X|\nabla^m\alpha|^p\,d\text{vol}\,.
$$
For such forms we say that $d\alpha=\beta$ ($\alpha$ a $k$-form) in
the sense of distributions provided
$$
\int_X\beta\wedge\gamma+(-1)^k\alpha\wedge d\gamma=0\,,
$$
for all compactly supported smooth $(n-k)$-forms $\gamma$.

For the wedge product and pull-back we take the most concrete
definition possible, namely a coordinate definition. There are
fancier ways to address this issues used in geometric measure
theory, but we will not need them in this paper. The coordinate
definition of pull-back
\begin{align*}
u^*&\left(\alpha_{k_1,\dots,k_p}(x^1,\cdots,x^n)dx^{k_1}\wedge\dots
dx^{k_p}\right)\\& = 
\sum_\ell\alpha_{k_1,\cdots,k_p}(u^1(y),\dots,u^n(y))\;\;\frac{\partial
u^{k_1}}{\partial y^{\ell_1}}\cdots\frac{\partial u^{k_p}}
{\partial
y^{\ell_p}}
\;\;dy^{\ell_1}\wedge\cdots dy^{\ell_p}\,,
\end{align*}
requires that the function $u$ have sufficient regularity (but $u$
need not be smooth) and one must prove that it is well-defined
(independent of coordinate system) for sufficiently regular $u$.
Similar comments hold for the coordinate definition of the wedge
product.

To define manifold-valued Sobolev spaces,
the target manifold, $Y$, is assumed to be isometrically embedded
into some Euclidean space ${\mathbb R}^N$, and $W^{s,p}(X,Y)$
is understood as follows:
$$
W^{s,p}(X,Y)=\{u\in W^{s,p}(X,{\mathbb R}^N)|\,u(x)\in Y\;\hbox{for a.e.}\;x \}.
$$
We will include several useful examples of functions belonging to
some of these Sobolev spaces after we review the definition of the
Faddeev and Skyrme functionals.

Over the past decade there has been a considerable progress made in
understanding the topology of Sobolev, $W^{s,p}$, maps between
manifolds. The list of publications on the subject is large and ever
growing. An excellent overview of the main directions in this
research is given by H. Brezis in \cite{Brezis}. A useful
observation is that when the Sobolev space does respect the
conventional topology, the conventional topological properties can
be extended to the discontinuous Sobolev maps. When the Sobolev
space does  not respect the conventional topology, various bad
things happen (for example, the homotopy classes of such maps
degenerate: every map can be deformed to a constant). Of course, one
should make clear what `respect' means.

If we talk about homotopy invariants of smooth maps from $M$ into
$N$, and we have analytic expressions (of course, one should look
for least restrictive analytic expressions) for all those
invariants, and we see that the expressions make sense (can be
computed) for {\it all} maps in some Sobolev space $W^{s,p}(M;\,N)$,
then we say that $W^{s,p}(M;\,N)$ respects the (conventional)
topology. Then we would expect that the numerical range of the
invariants computed for maps in $W^{s,p}(M;\,N)$ is the same as for
smooth maps (and attempt to prove this). Then we would expect that
the homotopy classes for $W^{s,p}$-maps each contains the
corresponding class of smooth maps (and attempt to prove this). On
the other hand, if the space $W^{s,p}(M;\,N)$ does not respect the
topology, then this nice picture most probably breaks down, and we
may study where it breaks down and to what extent. This general
point of view has been and is guiding our work on the topology of
Skyrme and Faddeev models. We bring in an additional point that it
is important to consider other functionals defined on the maps $M\to
N$, and interesting functionals come from physics.


\subsection{The Faddeev and Skyrme functionals}\label{fadsky}

The Skyrme functional was introduced in 1961 as an effective field
theory for nucleon interactions. This model received renewed
interest in 1983 when Witten described one Fermionic quantization of
the model with structure group SU$(3)$ \cite{witten}. The functional
was originally formulated for maps from $\R^3$ to SU$(2)$. The
natural generalization defined on maps from a $3$-manifold $M$ to a
Lie group $G$ is
$$
E(u) \,=\,\int_M|u^{-1}du|^2\,+\,| u^{-1}du\wedge u^{-1}du|^2\,d\hbox{vol}_M\,.
$$
The original Faddeev functional was defined for maps from $\R^3$ to
$S^2$. It extends naturally to maps $\varphi:M\to S^2$,
\cite{Faddeev}. \be\label{faddeev} E(\varphi)\,=\,\int_M
|d\varphi|^2\,+\,|d\varphi\wedge d\varphi|^2\,d\hbox{vol}_M. \ee
Just to be clear about what is meant here, writing
$\varphi=\varphi^1{\bf i}+\varphi^2{\bf j}+\varphi^3{\bf k}$, we
have
$$
|d\varphi|^2\,=\,|d\varphi^1|^2+|d\varphi^2|^2+|d\varphi^3|^2\,,
$$
$$
|d\varphi\wedge d\varphi|^2\,=\, |d\varphi^1\wedge d\varphi^2|^2+|d\varphi^2\wedge d\varphi^3|^2+|d\varphi^3\wedge d\varphi^1|^2\,.
$$
The lowest regularity of admissible maps for the Faddeev functional
is a little bit better than $W^{1,2}(M; S^2)$, for the integral of
the quartic term to be finite.

Analysis of Sobolev maps is the analysis of the functional on the
space of maps that corresponds to the Sobolev norm. Some of the
Sobolev norms preserve homotopy information, some do not. One of the
points we would like to make is that there are functionals other
than Sobolev norms, that may be as interesting or even more
interesting  from the viewpoint of the topology of manifolds or from
the viewpoint of Physics. The Skyrme and Faddeev functionals are two
such examples. Of course preserving homotopy is desirable from the
viewpoint of topology, see \cite{bmss} for the reasons why these
functionals are `natural' from the viewpoint of Physics.


\subsection{Sample functions}\label{sample}

Good functions to keep in mind when thinking about
Sobolev spaces are radial functions with variable  rates of growth
and oscillation. 
Let 
$D^n$ denote an open unit disk (ball) in $\R^n$ centered at the origin. The function 
$f:\;D^n\to \R$ given by
$f(x)=|x|^{-n/q}$ is in $W^{s,p}(D^n)$ with  $s>0$ if $|x|^{-(n/q + s)}$ is in $L^p$, which means 
$(n/q + s)\,p - (n-1) < 1$, i.e., $n/q + s < n/p$. 
The function $|x|^{-n/q}$ does not belong to $L^q(D^n)$. 
Note that, by the 
Sobolev embedding theorem, the space $W^{s,p}(D^n)$ with 
$s>0$ embeds continuously into
$L^q(D^n)$ when the indices satisfy 
$1\le p,\, q$ and $s - n/p\ge - n/q$.

\medskip
\noindent{\bf Example 4.}\label{eg4} 
Arguably the most useful map from $D^k$ to $\R^k$ is the map 
$\phi:\,x \mapsto x/|x|$. This map is bounded and its derivative behaves like 
$1/|x|$ which is in $L^p(D^k)$ provided $p < k$. Thus, 
$\phi\in W^{1,p}(D^k,\,\R^k)$ for $1\le p < k$. 
This map can be generalized to higher dimensions 
as a map 
$\phi:
D^k\times D^{n-k} \to \R^n$ given by
$\phi(x,y)=(x/|x|,y)$. It follows that $\phi\in
W^{1,p}(D^k\times D^{n-k}, \R^n)$ if and only if $p<k$.
We
conclude that there is no reasonable sense in which a $W^{1,p}$ map
will preserve homotopy information on the $k$-skeleton for $p<k$.
This illustrates the result of White \cite{White}: The homotopy
classes of maps with one derivative in $L^p$ restricted to the
$k$-skeleton of a manifold agree with the homotopy classes of smooth
maps provided $p$ is larger than $k$. In light of this, when $M$ is a
$3$-manifold a `natural' Sobolev space of maps 
$M\to S^2$ to
consider is $W^{1,3}(M; S^2)$. This is because the degree and many
other homotopy invariants can be defined for such functions. Brezis
and Nirenberg \cite{BN} showed that the degree can be defined for functions
belonging to VMO -- the space of functions of {\it vanishing mean
oscillation}. This is a larger space than $W^{1,3}(M; S^2)$. By
definition a function $f$ is in VMO if
\[
\lim\limits_{\epsilon\to 0}\;\;\sup\limits_{D_R,\,0<R<\epsilon}
|D_R|^{-1}\int_{D_R}|f-|D_R|^{-1}\int_{D_R}f|=0\,,
\]
where $D_R$ is a disk of radius $R$. Functions in VMO (and thus
$W^{1,3}(M; S^2)$) can be approximated arbitrarily closely in the
BMO norm by smooth functions, see \cite{BN}. As we show in this
paper the functions with finite Faddeev energy are also `natural'
from the point of view of topology. There are finite Faddeev energy
functions that are not in VMO, so arguments in the finite energy
case are more subtle; see Example 5 below.

Several maps similar to the ones in Example 4 may be used to clarify
the distinction between $W^{1,p}$ maps and finite Faddeev or Skyrme
energy maps. The map $\eta_1:D^3\to S^2\subset \R^3$ given by
$\eta_1(x)=x/|x|$ is in $W^{1,p}$ for $p<3$, but is not in
$W^{1,3}$ and does not have finite Faddeev energy. The composition
of the projection of $S^3$ to $D^3$ with this map 
$\eta_1$ can be
patched into any smooth map from a $3$-manifold to obtain a similar
example. It follows using our observation after Example 4 that our
results extending homotopy invariants to finite Faddeev energy or
$W^{1,3}$ maps are sharp.

\medskip
\noindent{\bf Example 5.}\label{eg5} The function, $\eta_2:D^3\to
S^2$ given by $\eta_2(x)=\cos(\ln|x|){\bf i}+\sin(\ln|x|){\bf j}$ is
in $W^{1,2}$ and has finite Faddeev energy but is not in $W^{1,3}$,
and the function $\eta_3:D^3\to S^1$ given by
$\eta_3(x)=\cos(\ln|\ln (|x|)|){\bf i} + 
\sin(\ln|\ln (|x|)|){\bf j}$ is in
$W^{1,3}$ but is not continuous.  These last two functions may be
patched into maps from an arbitrary $3$-manifold. Furthermore they
may be composed with maps into $S^3$ or any non-trivial compact Lie
group. These last examples show that the finite energy maps 
need not be continuous, so a special care is needed when defining
homotopy invariants for maps in these spaces. In fact, Diego
Maldonado pointed out that
 the function $\eta_2$ is not even in VMO. To see this recall
that a vector-valued function is in VMO exactly when each component
is in VMO and notice that
\[
\int r^2\cos(\ln r)\,dr=\frac{3}{10}r^3\cos(\ln
r)+\frac{1}{10}r^3\sin(\ln r) + C\,,
\]
so the integral in the definition of VMO can be evaluated explicitly
for disks centered at the origin.

The following example is a variant of Example 4 that 
illustrates certain issues that we resolve in this paper.

\medskip
\noindent{\bf Example 6.}\label{eg6} Let $\vartheta$ be coordinates
on $S^{n-1}$, so that generalized polar coordinates on $\R^n$ may be
written $(r,\vartheta)$. Using these coordinates for $\R^n$ and
generalized cylindrical coordinates $(r,\vartheta,z)$ for $\R^{n+1}$,
the stereographic parametrization of $S^n$ may be expressed as
$(r,\vartheta)\mapsto(\frac{2r}{r^2+1},\vartheta,\frac{r^2-1}{r^2+1})$.
In this parametrization the volume form on $S^n$ may be written as
$$
d\hbox{vol}_{S^n}=2^n(r^2+1)^{-n}d\hbox{vol}_{\R^n}=
2^nr^{n-1}(r^2+1)^{-n}\,dr\,d\hbox{vol}_{S^{n-1}}\,.
$$
Consider the map $\Phi:\,S^{n}\to S^{n}$ given in terms of coordinates 
in $\R^n$ by
$$
\Phi(r,\vartheta)=\left\{\begin{array}{lll}(r,\vartheta)&\hbox{if}&r\le 1\\
(1,\vartheta)&\hbox{if}&r\ge 1\end{array}\right.
$$
Geometrically the map $\Phi$ is the identity on
the lower hemisphere glued to the ubiquitous map $x\mapsto x/|x|$ 
on the the upper hemisphere ($\Phi$ projects points in the upper hemisphere to
the point on the equator meeting the great circle passing through
the given point and the north pole). The derivatives of $\Phi$ 
are bounded on the lower hemisphere and grow as $r$ on the upper 
hemisphere. For the integral 
\[
\int^\infty {r^{n-1}\over (1 + r^2)^n} \;r^p\;dr
\]
to be finite, one must have $p < n$. This implies 
$\Phi\in W^{1,p}(S^n,S^n)$ for $p < n$. 
This is slightly less regularity than what is required for a
well-defined degree.  However, the integral expression 
for the degree of $\Phi$ does make sense and  
yields $\int_{S^n}\Phi^*\omega_{S^n}=\frac12$. This example can
easily be modified to show that every real number is the `degree' of
a map in $W^{1,p}(S^n,S^n)$ when $p<n$.

\medskip
\noindent{\bf Example 7.}\label{eg7} The usual identities from
differential topology no longer hold when the regularity of the
functions in question is too low. In particular, 
the pull-back and the differential may not commute. 
 Consider the ubiquitous map $u;\,z\mapsto z/|z|$ 
from $D^2$ to  
$\R^2$.  If $(t,\phi)$ and
$(r,\theta)$ are polar coordinates on the domain 
and the target  respectively, we have  
$u(t,\phi)=(1,\phi)$. This is clearly in $W^{1,p}$ for every $p<2$.
Set $\alpha=r^2\,d\theta$, so that $d\alpha=2r\,dr\wedge d\theta$.
It follows from the definition that 
$u^*(d\alpha)=0$.  On the other hand 
$u^*\alpha = \,d\phi$ and 
$d(u^*\alpha)$ is a multiple of the delta-function:
\[
\left( f\,,d(u^*\alpha)\right) = - \int_{D^2} df\wedge u^*\alpha = - \lim\limits_{\epsilon\to 0} 
\int_{D^2\setminus \{|x|<\epsilon\}}df\wedge d\phi = 
\lim\limits_{\epsilon\to 0}\int_{|x| = \epsilon} f\,d\phi = 
{2\,\pi}\,f(0)\,, 
\] 
for any test function $f$ in $D^2$. This example generalizes to show that the pull-back with respect to
the maps from Example 4 does not commute with exterior
differentiation. As we have remarked, there is such a map in
$W^{1,p}(D^n,D^n)$ for every $p<n$. 

The maps whose pull-backs commute with the differentials of forms of certain degrees have special regularity properties and have been studied, 
see for example \cite[Chapter I.3]{GMS2}. 
In our present paper this property has a prominent role. We need the intertwining maps $u:\,M^3\to S^3$  
to have it in order to prove that their degree 
is an integer. Starting from Proposition \ref{rep} 
we prove this property again and again for various maps. We cannot apply the machinery developed in 
\cite{GMS2} because we work not in the Sobolev  
but rather in finite energy environment. We develop 
our own technique to handle this situation.

\section{Local representations}\label{sec1}

Two local representation theorems are
key technical tools used in this paper. The first is the following result
about the local structure of flat connections that we
proved in \cite[Lemma 3]{AK} and called the nonlinear Poincar\'e
lemma. Let $G$ be a compact Lie group and
$\mathfrak g$ be its algebra Lie. Let $I^m$ denote a unit cube
in $\mathbb R^m$.
\begin{lemma}\label{poi}
Given any $L^2$ $\mathfrak g$-valued $1$-form $A$ on $I^3$ such that
\be\label{flat} dA\,+\,\frac12\,[A,\,A]\,=\,0 \ee in the sense of
distributions, there exists $\,u\in W^{1,2}(I^3,\,G)\,$ such that
$\,u^{-1}\in W^{1,2}(I^3,\,G)\,$ and $\,A\,=\,u^{-1}\,du$.
Furthermore, for any two such maps, $\,u\,$ and $\,v$, there exists
$\,g\in G\,$ so that  $\,u(x)\,=\,g\cdot v(x)$, for almost every
$\,x\in I^3$. If $A\in W^{k, 2}$ then we can take $u\in W^{k+1, 2}$.
If $A\in C^\infty$ then we can take $u\in C^\infty$.
\end{lemma}

\noindent{\bf Proof.}\ \ The case $A\in L^2$ is 
proved in \cite[Lemma 3]{AK}. If we know that $A\in W^{1,2}$, the map $u$ constructed in \cite[Lemma 3]{AK} will belong to $W^{2,2}$. Indeed, differentiating $\partial_j u = u\,A_j$, we obtain 
$\partial_k\partial_j u = u\,A_k\,A_j + u\,\partial_k A_j$. Because $u$ is bounded 
as a map into a compact group, the norm $\|u\,A_k\,A_j\|_{L^2}$ is less than a constant times $ \|A\|_{L^4}^2$, which is finite 
due to 
Sobolev embedding $W^{1,2}\subset L^4$ in dimension $3$. If $A$ has derivatives of order $2$ or higher in $L^2$, then a similar argument shows that $u$ has 
one derivative more than $A$ in $L^2$. \hfill$\square$%

\begin{remark}
The non-linear Poincar\'e lemma also holds for flat connections
defined on any domain that is bilipschitz equivalent to the unit
cube.  Assuming that $\Omega$ is a domain bilipschitz equivalent to
a cube and $A$ is a flat connection in $L^2(\Omega)$, we can
pull-back $A$ to the cube and the result will still be flat (as one
can see by the change of variables formula), so we can apply the
non-linear Poincar\'e lemma to obtain the local developing map in
$W^{1,2}(I^m)$ and change coordinates to obtain the desired local
developing map on the domain. The change of variables formula
implies that $f\circ g$ is in $W^{1,p}(X)$ exactly when $f$ is in
$W^{1,p}(Y)$ provided $p\ge 1$ and $g:X\to Y$ is bilipschitz.
\end{remark}
\begin{remark}\label{triang}
In the historical development of differential topology the existence
and uniqueness of piecewise linear structures on smooth manifolds
was a fundamental question. In order to interpolate between the
smooth and piecewise linear categories, the notion of piecewise
smooth functions was introduced.  Recall that an abstract simplicial
complex $K$ is a collection of finite non-empty subsets of a fixed
vertex set such that any non-empty subset of an element of the
collection is in the collection and the union of the entire
colleciton is the vertex set. There is a standard construction of a
geometric realization of an abstract simplicial complex (denoted by
$|K|$) obtained by associating one standard basis vector for each
vertex and taking the union of the convex hulls of the points
associated to each set in the collection. The convex hull of the
points associated to one of the sets in the collection is called a
geometric simplex. A map from the geometric realization of an
abstract complex into a manifold is called piecewise smooth if the
restriction of the map to each geometric simplex is equal to the
restriction of a smooth map from an open set containing the simplex.
The closed star of any point in the geometric complex (denoted by
st$(p)$ is the union of all geometric simplices containing that
point. One defines a differential of a piecewise smooth map
$f:|K|\to M$ at a point as the map $df_p:\text{st}(p)\to T_{f(p)}M$
given by $df_p(v):=\frac{d}{dt}f(tv+(1-t)p)|_{t=0}$. A map from a
geometric complex to a subset of a manifold is a piecewise smooth
equivalence if it is piecewise smooth, bijective with bijective
differentials. It is a classical fact due to Whitehead that every
smooth manifold is piecewise smoothly equivalent to some geometric
complex; see \cite{Munkres}. In addition the image of any
intersection of closed stars of vertices is piecewise smoothly
equivalent to the unit cube. We use these facts when we need to
decompose our manifold into reasonable domains later. Notice that
piecewise smooth equivalence implies bilipschitz equivalence. From
here forward we will use $\Omega$ to denote a domain piecewise
smoothly equivalent to a cube.
\end{remark}

We now describe the second local representation theorem. This
theorem states that any sufficiently regular map $\vp:\,\Omega\to
S^2$ may be written in the form $\vp=u^{-1}{\bf i}u$ for some lift
$u:\Omega\to S^3$. Notice that this result must be local because
there can be no lift $u:M\to S^3$ of a map $\vp:M\to S^2$  
when $\vp^*\mu_{S^2}\ne 0$ because $H^2(S^3)=0$. Also
notice that when a lift does exist, there should be many lifts
because $S^3$ has one more dimension than $S^2$. In fact there is an
infinite-dimensional group of gauge symmetries that act on the
possible lifts. One may impose a local gauge-fixing condition to
remove this ambiguity. As usual requiring the slice to be
perpendicular to the orbit of the gauge group is a reasonable choice
for gauge-fixing condition. This is the interpretation of the gauge
fixing condition (\ref{gaugecond}) that we include in our local
representation theorem.

\subsection{Finite energy pull-back of the area form}

Before we get to the proof of Proposition \ref{rep} we would like to
point out once again the main technical difficulties one encounters when
generalizing usual results from differential topology to functions
with low regularity/summability. 
\bigskip

\noindent$\bullet$\ \   
The pull-back and the differential do not necessarily commute 
(Example 7 from Section \ref{sample}).  
To establish $df^*\alpha=f^*d\alpha$ we need to fully exploit the 
structure of the form $\alpha$ and special properties (e.g.  
boundedness) of the map $f$. Where na\"ive mollification does not 
work, we decompose the form and approximate appropriate parts 
in a way preserving cancellations, 
apply the formula and take limits. 
\bigskip

\noindent$\bullet$\ \  
The product formula 
$d(\alpha\wedge\beta)=d\alpha\wedge\beta+(-1)^{|\alpha|}\alpha\wedge
d\beta$ does not necessarily hold if $\alpha$ and/or $\beta$ 
lack summability. Again, we use the structure of the forms to show that 
$\alpha\wedge\beta$ and each
term on the right is in $L^1_{loc}$.

\bigskip

We need to establish an auxiliary result before considering the
second local representation theorem. Namely, the pull-back of the
normalized volume form $\omega_{S^2}$ by a finite Faddeev energy map
$\vp$ is a closed 2-form on $\Omega$.
\begin{remark}
In their paper \cite{LY}, Fanghua Lin and Yisong Yang {\it include}
the requirement of the (distributional) closedness of the pull-back
in the definition of their class $X_2$ of maps where minimizers of
the Faddeev energy are sought. Our lemma below shows that finite
energy implies $d\vp^*\omega_{S^2} = 0 $.
\end{remark}

\begin{lemma}\label{areaS^2} Let $\Omega$ be a domain piecewise smoothly
equivalent to the cube. If $\vp$ is a finite Faddeev energy map in
$W^{1,2}(\Omega,S^2)$, then the pull-back, $\vp^*\omega_{S^2}$, is
closed: $d\vp^*\omega_{S^2} = 0$ in the sense of distributions. In
particular, 
\be\label{dvp^3} 
d\vp\wedge d\vp\wedge d\vp\,=\,0\,. 
\ee
\end{lemma}

\begin{remark}\label{R3s2area}
The same result holds for finite Faddeev energy maps on $\R^3$.
Indeed to to check that $\langle\vp^*\omega_{S^2},d\phi\rangle=0$
for a smooth test form $\phi$ it is sufficient to check it on the
closure of the support of $\phi$. This is certainly contained in
some cube and the restriction of $\vp$ to this large cube satisfies
the hypothesis of the theorem.
\end{remark}
\noindent{\bf Proof.} Recall from Lemma \ref{normform} that,
\[
\vp^*\omega_{S^2}\,=\,-\,\frac{1}{8\pi}\,\vp\,d\vp\wedge d\vp\,;
\]
in particular the form $\vp\,d\vp\wedge d\vp$ is real-valued.
Consider $y^{-1}\,dy$ on $\mathbb H - \{0\}$ and set $A =
\vp^*(y^{-1}\,dy) = - \vp\,d\vp$. By mollification we can find a
sequence of functions $\vp_n:\Omega\to \mathbb{H}$ such that
$\vp_n\to \vp$ in $W^{1,2}$. Applying the product rule to
$\vp_n\,d\vp_n$ and passing to the limit we see that
\[
dA\,=\,-\,A\wedge A\,=\,-\,d\vp\wedge d\vp
\]
in the sense of distributions. Thus,
\[
-\,\vp\,d\vp\wedge d\vp\,=\,\vp\,dA\,.
\]
Since $\vp$ is in $W^{1,2}(\Omega,S^2)$ and has finite Faddeev
energy, we conclude that $A\in L^2$ and $dA\in L^2$. Let
$T_\epsilon$ denote a mollification operator. Then
\[
d\left( \vp_n\,d T_\epsilon(A)\right) = d\vp_n\wedge d T_\epsilon(A)
= d\vp_n\wedge T_\epsilon(d A),
\]
which converges to $d\vp\wedge dA$ in $L^1$. Thus, in the sense of
distributions,
\[
d\vp^*\omega_{S^2}\,=\,-\,{1\over 8\pi}\,d\vp\wedge d\vp\wedge
d\vp\,.
\]
Now, check that
\[
d\vp\,=\,-\,\frac12\,[A,\,\vp]\,
\]
and compute \be\label{triple}
\begin{aligned}
8\,d\vp\wedge d\vp\wedge d\vp \,= & - [A,\vp]\wedge
[A,\vp]\wedge [A,\vp] \\
=&\,-\,\hbox{Re}\,\left( A\,\vp\wedge A\,\vp\wedge A\,\vp \right.
+\; A\,\vp\wedge A\wedge A \\
&- \vp\,A\wedge A\,\vp\wedge A\,\vp  - \;\vp\,A\wedge A\wedge A\\
&+ A\wedge A\wedge A\,\vp  - \;A\wedge A\wedge \vp\,A \\
&+ \vp\,A\wedge \vp\,A\wedge A\,\vp \left. - \;\vp\,A\wedge
\vp\,A\wedge \vp\,A\right)\,.
\end{aligned}
\ee
Using the fact that factors can be cycled under the real part we see that
all terms happily  cancel out.   \ \hfill $\square$%
\bigskip

\subsection{Representation of $S^2$-valued maps on cubes}

We are now in a position to prove the local representation theorem
for $S^2$-valued maps.

\begin{proposition}\label{rep} 
Let $\Omega$ be a bounded region in $\R^3$ piecewise smoothly
equivalent to the cube. If $\varphi :\,\Omega\to S^2$ is a finite Faddeev energy map, 
then there is a map $u\in
W^{1,2}(\Omega,S^3)$ so that $\varphi=u^{-1}\,{\bf i}\,u$. In
addition, $u^{-1}du\wedge u^{-1}du \wedge u^{-1}du\in L^{3/2}(\Omega)$, and
$u$ has the following important property: For any smooth
$\R^3$-valued function $f$ on $S^3$, 
\be\label{u*d=du*}
u^*\left(d\langle f,\,y^{-1}\,dy\wedge y^{-1}\,dy\rangle
\right)\,=\, d\left(u^*\left(\langle f,\,y^{-1}\,dy\wedge
y^{-1}\,dy\rangle \right)\right)\,. 
\ee 
Here $y^{-1}dy$ is the usual
Maurer-Cartan form on $S^3$. Furthermore, for any two such maps $u$
and $v$ there is a map $\lambda$ in 
$W^{1,2}(\Omega,S^1)$ so that
$v=\lambda\, u$. Given a smooth metric on $\Omega$ one can choose
$u$ as above with 
\be\label{gaugecond}
\delta\langle
u^{-1}du,\vp\rangle=0,\quad \text{and} \quad i^**(\langle
u^{-1}du,\vp\rangle)=0\,,
\ee 
where $\delta$ is the codifferential
defined via the metric, $*$ is the Hodge star operator and
$i:\partial \Omega\to \Omega$ is the inclusion. Such a lift is
unique up to left multiplication by a unit complex number. A similar
result also holds for maps in $\vp\in W^{1,3}(\Omega,\,S^2)$. In this case
$u^{-1}du\wedge u^{-1}du \wedge u^{-1}du\in L^{1}$, and
(\ref{u*d=du*}) holds.
\end{proposition}

\begin{remark}\label{cart1} That $u\in W^{1,2}$,
$u^{-1}du\wedge u^{-1}du \wedge u^{-1}du\in L^1$, and equality
(\ref{u*d=du*}) holds imply that the map $u$ is cartesian,
$u\in\hbox{cart}^1(\Omega,\,\mathbb R^4)$ in the sense of
\cite[Section I.3.2]{GMS2}.
\end{remark}

\smallskip\noindent{\bf Proof.}
Assume for a moment that we knew that such a map $u$ existed. Then a
computation starting with the derivative of $\varphi=u^{-1}\,{\bf
i}\,u$ leads to $\varphi^{-1}d\varphi=a+\varphi \,a\,\varphi$ where
$a=u^{-1}du$. Since $\varphi^{-1}d\varphi$ is perpendicular to
$\varphi$, we can solve this last equation for $a$ by splitting it
into directions parallel and perpendicular to $\varphi$. We see that
this equation holds exactly when $a=\frac12\, \varphi^{-1}\,d\varphi
+\varphi\,\xi$ for some real valued $1$-form $\xi$. In fact $\xi$ is
just the $\vp$ component of $a$. Since $a$ is flat we would have
$0=da+a\wedge a=\varphi\, d\xi-\frac14 \, d\varphi\wedge d\varphi$
or equivalently $d\xi = -\frac14 \,\varphi \,d\varphi\wedge
d\varphi$. We will turn this around by solving for $\xi$ then $a$
and finally $u$.

We already saw in the proof of Lemma \ref{normform} that the form
$\varphi\, d\varphi\wedge d\varphi$ is real. That it is closed
follows from Lemma \ref{areaS^2}. When $\varphi\in W^{1,3}$ this
form is in $L^{3/2}$ and when $\varphi$ has finite Faddeev
energy this form is in $L^{2}$. It follows from the usual Poincar\'e
lemma  that there is a $\xi$ in $W^{1, 3/2}$ or 
$W^{1,2}$
respectively so that $d\xi = -\frac14 \,\varphi\, d\varphi\wedge
d\varphi$, \cite{Morrey}. In fact there is a unique such form
satisfying 
\be\label{xi2} 
d\xi = -\frac14 \,\varphi\, d\varphi\wedge
d\varphi\,,\quad \delta\,\xi = 0\,,\quad\hbox{in}\quad \Omega,\qquad
i^**(\xi) = 0\,. 
\ee  
Set $a=\frac12 \,\varphi^{-1}d\varphi
+\varphi\,\xi$ and observe that the Sobolev embedding theorem
implies that $a$ is in $L^2$ in either case. Now, 
taking into account  (\ref{xi2}) and 
that pointwise $\vp$ is purely imaginary 
and $\vp^2 = -1$, we obtain
\[
da = - \frac12\,\vp^{-1} d\vp\wedge \vp^{-1} d\vp 
+ d\vp\wedge \xi + \vp\,d\xi = 
- \frac14\,d\vp\wedge d\vp + d\vp\wedge \xi\,.
\]
For the same reasons,
\[
a\wedge a = \frac14\,\vp^{-1} d\vp\wedge \vp^{-1} d\vp 
+ \frac12\,\vp^{-1} d\vp\wedge \vp\,\xi 
+ \frac12\,\vp\,\xi\wedge \vp^{-1} d\vp = 
\frac14\,d\vp\wedge d\vp - d\vp\wedge \xi\,.
\]
Thus, $da + a\wedge a = 0$, i.e., 
$a$ is flat. The
nonlinear Poincar\'e lemma (Lemma \ref{poi}) now implies the
existence of a $w\in W^{1,2}(\Omega;\,S^3)$ with 
$a=w^{-1}dw$.

Compute: $(w\varphi w^{-1})^{-1}d(w\varphi
w^{-1})=w\left(\varphi^{-1}d\varphi - A -\varphi\,
A\,\varphi\right)w^{-1}=0$. Since $\vp(x)\in S^2$ for almost all
$x$, the product $w\,\varphi\, w^{-1}$ takes values in $S^2$ almost
everywhere as well. Since its derivative vanishes, we conclude that
there is a $y\in S^2$ and thus a unit quaternion $p\in S^3$ so that
$w(x)\,\varphi(x)\, w(x)^{-1}=y=p^{-1}{\bf i}\,p$ almost everywhere.
Clearly $u=p\,w\in W^{1,2}$ satisfies $\varphi = u^{-1}{\bf i}\,u$.
When $\varphi\in W^{1,3}$ we have $u^{-1}du = a\in L^3$, thus
$\hbox{Re}(u^{-1}du\wedge u^{-1}du\wedge u^{-1}du)\in L^1$.

Now assume that we only know that $\varphi\in W^{1,2}$ and has finite
Faddeev energy. Recall that
\[
u^{-1}du = a = \frac12 \,\varphi^{-1}d\varphi +\varphi\,\xi \,,
\]
where $\xi$ is a solution of the elliptic problem (\ref{xi2}). Clearly, 
$\xi=\langle u^{-1}du,\vp\rangle$, and so  the gauge fixing
conditions (\ref{gaugecond}) are satisfied. As we have seen, 
\[
a\wedge a = \frac14\,d\vp\wedge d\vp - d\vp\wedge \xi\,.
\]
Since $\vp$ is assumed to have finite Faddeev energy, $d\vp\in L^2$ and $d\vp\wedge d\vp\in L^2$. Also, the solution of problem (\ref{xi2}) satisfies $\xi\in
W^{1,2}\subset L^6$. Thus, $a\wedge a\in L^{3/2}$. By the isoperimetric
inequality (for a parallelopiped in $\mathbb R^3$)
\[
|a\wedge a\wedge a|\,\le\,C\,|a\wedge a|^{3/2}\,,
\]
thus  $a\wedge a\wedge a$ is in $L^1$. We will see a second proof
establishing that $a\wedge a\wedge a\in L^{3/2}$, independently of
the isoperimetric inequality later.

Given two maps $u$ and $v$ with $u^{-1}\,{\bf i}\,u = v^{-1}\,{\bf
i}\,v$ set $\lambda = v\,u^{-1}$ and notice that $v\,u^{-1}{\bf
i}={\bf i}\,v\,u^{-1}$ implies that $\lambda$ takes values in $S^1$
hence lives in $W^{1,2}(\Omega,S^1)$. We can write $\lambda$ in the
form $\lambda(x) = \exp 2\pi\,{\bf i}\,\theta(x)$ for some
real-valued function $\theta\in W^{1,2}(U,\,\mathbb R)$. This
follows from \cite[Lemma 1]{BZ}, or from our non-linear Poincar\'e
Lemma \ref{poi} after noticing that $\lambda^{-1} d\lambda$ is a
flat, ${\bf i}\mathbb R$-valued connection. In any case, $v^{-1} dv
= u^{-1} du + \vp\,2\pi\,d\theta$, and assuming that both $u$ and
$v$ satisfy the gauge fixing condition (\ref{gaugecond}) implies
that $\delta d\theta=0$, so we must have $d\theta = 0$, i.e.,
$\theta(x) = const$.

It remains to prove formula (\ref{u*d=du*}). We have
\[
u^*\left(\langle f,\,y^{-1}dy\wedge y^{-1}dy\rangle \right)\,=\,
\langle u^*f, a\wedge a\rangle\,.
\]
Thus our goal is to prove
\begin{equation}\label{goal}
d\left(\langle u^*f, a\wedge a\rangle\right)\, =\,u^*d\langle f,
y^{-1}dy\wedge y^{-1}dy\rangle.
\end{equation}
When $u\in W^{1,3}$ this is a straight-forward application of the
approximation argument. However when $u$ is only assumed to have
finite energy the argument is not obvious at all. The difficulty here,
at the surface of it, is that $a\wedge a$ may not be in $L^2$, and
the best we can say about $u^*df$ is that it is in $L^2$. This means
that it is hard to see that the approximations  converge in $L^1$ to
the right-hand side  to make the approximation argument work. Thus,
the only hope is to look at the structure of these expressions and
find the right cancelations to get the proper convergence.
\medskip

We start by re-writing the right-hand side of 
(\ref{goal}). To shorten the formulas it is convenient to adopt the summation over the repeated 
indices rule and to introduce the following temporary notation. 

Set ${\bf e}_1 = {\bf i}$, ${\bf e}_2 = {\bf j}$, 
and ${\bf e}_3 = {\bf k}$. Let $\theta$ denote 
the Maurer-Cartan form $y^{-1}dy$ and write 
\[
y^{-1}dy = \theta = \theta^i\,{\bf e}_i\,.
\]
The real $1$-forms $\theta^1$, $\theta^2$, and $\theta^3$ 
form an orthonormal basis in $T^*S^3$. Denote by 
$X_1$, $X_2$, $X_3$ the dual basis of $TS^3$,  
$\theta^m(X_\ell) = \delta_{\ell m}$. One easily checks that, for any function $f$,  
\be\label{df}
df=X_k(f)\theta^k.
\ee
We also have 
\[
\theta\wedge \theta = 2\,\left(
\theta^2\wedge\theta^3\,{\bf e}_1 + 
\theta^3\wedge\theta^1\,{\bf e}_2 + 
\theta^1\wedge\theta^2\,{\bf e}_3 \right) 
\]
and 
\[
\langle \theta,\,\theta\wedge \theta\rangle = 
- \,\theta\wedge \theta\wedge \theta = 6\,\theta^1\wedge \theta^2\wedge \theta^3\,.
\]
Writing an $\R^3$-valued (i.e., purely imaginary) function $f$ as $f=f^i{\bf e}_i$, we have
\[
d\langle f, \theta\wedge \theta \rangle =
\frac13\;X_m(f^m)\;\langle \theta,\,\theta\wedge \theta\rangle\,.
\]
We used the fact that terms such as 
$\langle X_2(f^1)\theta^2{\bf e}_1,\theta\wedge \theta \rangle 
= 2\,X_2(f^1)\theta^2\wedge\theta^2\wedge\theta^3$  vanish. Since  
\[
u^*\theta = u^{-1} du = a\,,
\]
the right-hand side of our goal (\ref{goal}) is
\begin{equation}\label{goalrhs}
u^*\left(d\langle f, \theta\wedge \theta  \rangle\right) =
\frac13\;u^*(X_m(f^m))\;\langle a,\,a\wedge a \rangle\,.
\end{equation}
 
Turning to the left-hand side of (\ref{goal}),  recall that 
\[
a = -\frac12 \,\varphi \,d\varphi +\varphi\,\xi
\]
with $\vp, \xi\in W^{1,2}$. To see and  
exploit cancellations, we use a special 
approximation $a_n$ to $a$; namely, 
\[
a_n := -\,\frac12\,\vp \,d\vp + \vp\,\xi_n\,,
\]
where $\xi_n$ is a sequence of smooth real $1$-forms 
converging to $\xi$ in $W^{1,2}$ (and hence in $L^6$). 
This gives
\[
a_n\wedge a_n = \frac14\,d\vp\wedge d\vp - d\vp\wedge \xi_n
\]
and one sees that $a_n\wedge a_n\to a\wedge a$ in $L^{3/2}$. It
follows that $\langle u^*f, a_n\wedge a_n\rangle \to \langle u^*f,
a\wedge a\rangle$ in $L^{3/2}$ (since $f$ is bounded), so
\[
d\langle u^*f, a_n\wedge a_n\rangle \to d\langle u^*f, a\wedge
a\rangle
\] 
as distributions.
We will see that this sequence also converges to $\,u^*d\langle f,
y^{-1}dy\wedge y^{-1}dy\rangle$ to establish our goal (\ref{goal}).

Note that $a_n\wedge a_n$ is in $L^2$ since $\vp$ has finite
Faddeev energy. Also, $d\left(a_n\wedge a_n\right)
\,=\, d\vp\wedge d\xi_n$ in the sense of distributions. 
Moreover, $d\vp\wedge d\xi_n \to d\vp\wedge d\xi$ 
in $L^1$, and $d\vp\wedge d\xi=\frac14\,\vp\,d\vp\wedge d\vp\wedge d\vp = 0$ by Lemma \ref{areaS^2}. Since 
$u^*f$ is bounded with derivative in $L^2$, we have  
\[
d\langle u^*f, a_n\wedge a_n\rangle= \langle d(u^*f), a_n\wedge
a_n\rangle + \langle u^*f, d\vp\wedge d\xi_n\rangle \,.
\] 
As $n\to\infty$, the second term on the right goes to $0$. Let us analyze the first term. 

Since $u$ is in $W^{1,2}$ and $f$ is smooth, 
$d(u^*f)=u^*(df)$. Applying  formula (\ref{df}) 
to $f = f^j\,{\bf e}_j$, we get
\[
d(u^*f)\,=\, u^*(X_i(f^j))u^*(\theta^i){\bf e}_j.
\]
Noting that $\theta^i = \langle \theta,\,{\bf e}_i\rangle$ and $a = u^*\theta$, we write  
\[
u^*(\theta^m)\,=\,
\langle a,\,{\bf e}_m\rangle \,=\,
\langle a - a_n,\,{\bf e}_m\rangle \,+\, \langle
a_n,\,{\bf e}_m\rangle\,.
\]
Then, 
\be\label{halfgoal}
\langle d\left(u^*f\right), a_n\wedge a_n\rangle =
u^*(X_i(f^j))\langle
a-a_n, {\bf e}_i\rangle\langle
a_n\wedge a_n, {\bf e}_j\rangle\,+\,
u^*(X_i(f^j))\langle
a_n, {\bf e}_i\rangle\langle a_n\wedge a_n, {\bf e}_j\rangle\,.
\ee
Since 
\[
a - a_n = \vp\,(\xi - \xi_n)\,,\qquad  
a_n\wedge a_n =
\frac14\,d\vp\wedge d\vp - d\vp\wedge \xi_n\,,
\] 
and $\xi_n\to \xi$ in $W^{1,2}$, and $\vp$ has finite Faddeev energy, and both $\vp$ and $u$ are bounded, 
the term 
\[
u^*(X_p(f^q))\langle a-a_n,e_p\rangle\langle a_n\wedge a_n,
e_q\rangle=u^*(X_p(f^q))\vp^p(\xi-\xi_n)\wedge \left(\frac14\,\langle
d\vp\wedge d\vp,e_q\rangle-d\vp^q\wedge \xi_n\right)
\]
converges to $0$ in $L^{3/2}$. Notice that this would not work if $a_n$ 
was simply a mollification of $a$ because $a_n$ would only approach $a$ in $L^2$ and $a_n\wedge a_n$ only approach $a\wedge a$ in $L^1$. 
With our approximation, the worst terms cancel.

Moving to the second term in (\ref{halfgoal}) notice that for any
$1$-form $b$ with values in $\mathbb{R}^3$ we can write
\begin{align*}
\langle b,\,{\bf e}_p\rangle&=b^p\\
b\wedge b & = 2\,b^1\wedge b^2 \;{\bf e}_3+\text{cyclic}\\
\langle b,\,b\wedge b\rangle&= 6\;b^1\wedge b^2\wedge b^3\\
\langle b,\, {\bf e}_p\rangle\,\langle b\wedge b,\, {\bf e}_q\rangle&= \frac13\,\langle b,
b\wedge b\rangle\;\delta_{pq}\,.
\end{align*}
Using this, the second term in (\ref{halfgoal}) becomes
\[
\frac13\,u^*(X_i(f^i))\langle a_n,a_n\wedge a_n\rangle\,.
\]
More cancelation takes place when we analyze this term. Indeed, 
\[
\langle a_n,\,a_n\wedge a_n\rangle\,=\,
\langle - \frac12\,\vp\,d\vp + \vp\,\xi_n,\,
\frac14\,d\vp\wedge d\vp - d\vp\wedge \xi_n 
\rangle = 
- \frac34\;\xi_n\wedge \vp\,d\vp\wedge d\vp\,.
\]
Now, $\langle a,\,a\wedge a\rangle\,=\,
- \frac34\;\xi\wedge \vp\,d\vp\wedge d\vp$. 
Since $\vp\,d\vp\wedge d\vp\in L^2$ and 
$\xi_n\to\xi$ in $L^6$, we conclude that 
\[
\langle a_n,\,a_n\wedge a_n\rangle \to 
\langle a,\,a\wedge a\rangle \qquad
\textrm{in}\quad L^{3/2}\,.
\]
Putting all parts together, with a slight abuse of 
notation we can write
\begin{align*}
\lim\limits_{n\to\infty} d\,\langle u^*f,\,a_n\wedge a_n\rangle & = 
\lim\limits_{n\to\infty} \langle d(u^*f),\,a_n\wedge a_n\rangle = \\ 
\lim\limits_{n\to\infty} 
\frac13\,u^*(X_i(f^i))\langle a_n,a_n\wedge a_n\rangle & = \frac13\,u^*(X_i(f^i))\langle a,\,a\wedge a\rangle\,.
\end{align*}
This agrees with our simplification (\ref{goalrhs}) of the
right-hand side of (\ref{goal}) and completes the argument.
Notice that $\langle a,a\wedge a\rangle=-\frac34\;\xi\wedge\vp\,d\vp
\wedge d\vp$ gives a second proof that $\langle a,a\wedge a\rangle$
is in $L^1$ (in fact $L^{3/2}$).

 \hfill$\square$

\begin{remark}
The decomposition $\langle a_n,a_n\wedge
a_n\rangle=-\frac34\;\xi_n\wedge\vp\,d\vp \wedge d\vp$ would hold for
any symmetric space, but it does not hold for homogeneous spaces
\cite{K}. Thus we expect that our argument would generalize to
symmetric spaces, but an alternate argument would be required for
homogeneous spaces.
\end{remark}

\noindent{\bf Example 8.} A $W^{1,2}$ lift of a finite Faddeev
energy map  need not have finite Skyrme energy. 
Indeed, let $\vp$ be a $W^{1,2}$ map into a great circle 
on $S^2$. Then $d\vp\wedge d\vp = 0$ and $\vp$ 
automatically has finite Faddeev
energy. Take a real, closed $1$-form 
$\xi\in L^2$ and set $a = - \frac12\;\vp\,d\vp + \vp\,\xi$. 
Clearly, $a\in L^2$. 
Because $d\xi = 0 = -\frac14\;\vp\,d\vp\wedge d\vp$, 
$a$ is flat. Thus, there exists a $W^{1,2}$ lift 
$u$ so that $u^{-1} du = a$.  As we have seen in the proof of Proposition \ref{rep}, 
$a\wedge a = \frac14\;d\vp\wedge d\vp - d\vp\wedge \xi$, which in the present case gives 
$u^{-1}du\wedge u^{-1}du = a\wedge a = - d\vp\wedge \xi$. Now it is easy to find $\vp$ and $\xi$ 
so that $d\vp\wedge \xi$ is not in $L^2$, and this 
would imply that $u$ has infinite Skyrme energy. 
At the same time, 
$\left(u^{-1}du\right)^{\wedge 3} = 0\in L^1$.  
\medskip

For $\Omega$ a neighborhood of the origin in $\R^3$, 
here is a concrete example of $\vp$ and $\xi$ 
that illustrates the above argument:
\[
\vp(x^1, x^2, x^3) = {\bf j}\,\exp\left({\bf i}\,(x^3)^{3/5}\right)\,,\quad 
\xi = r^{-6/5}\,dr\,.
\]
By inserting
smooth cut-off functions into the exponentials this example can be
patched into $\R^3$. Of course in this example we could have taken
$\xi=0$ and obtained a lift with finite Skyrme energy. It is an
interesting open question whether every finite Faddeev energy map
defined on a cube has a finite Skyrme energy lift.

\subsection{Representation of $S^2$-valued maps on $\R^3$}

We now establish a version of our local representation
proposition for maps $\vp:\R^3\to S^2$. 
We use it later to
give a correct proof that the Hopf invariant of a finite
Faddeev energy map from $\R^3$ to $S^2$ is an integer. 

The essential difference
between this and the previous case is that the domain is no longer
compact, so the 
proof of Proposition \ref{rep} does not automatically carry over to this case. The step where we use the Hodge decomposition is slightly modified. The Hodge decomposition theorem is usually stated for
compact manifolds, but it also holds when there is a suitable
balance between the function spaces and the geometry of the manifold
at infinity. The result that we need is presented below. 

\begin{lemma}\label{hodge}
Any $L^2$ form on $\R^3$ can be expressed 
in a unique way as
\[
\alpha=d\xi+\delta\omega\,
\]
with $\delta\xi = 0$, $d\omega = 0$, 
where $\xi$ and $\omega$ are in $L^6$, $d\xi$ and $\delta\omega$ are
in $L^2$.  
\end{lemma}
This result is well known to people working in mathematical hydrodynamics.

We now state and prove the representation result.

\begin{proposition}\label{R3rep}
If $\varphi$ is a finite Faddeev energy map from $\R^3$ to $S^2$
then there is a map $u:\R^3\to S^3$ with $du \in L^2$ so that
$\varphi=u^{-1}\,{\bf i}\,u$. In addition, $u^{-1}du\wedge u^{-1}du
\wedge u^{-1}du$ is in $L^1$ and $L^{3/2}$, and $u$ has the
following important property: For any smooth $\R^3$-valued function
$f$ on $S^3$, 
\be\label{u*d=du*2} 
u^*\left(d\langle
f,\,y^{-1}\,dy\wedge y^{-1}\,dy\rangle \right)\,=\,
d\left(u^*\left(\langle f,\,y^{-1}\,dy\wedge y^{-1}\,dy\rangle
\right)\right)\,. 
\ee 
Here $y^{-1}dy$ is the usual Maurer-Cartan
form on $S^3$. One also has
\[
u^*\left(d\langle
f,\,y^{-1}\,dy\wedge y^{-1}\,dy\rangle \right)\in L^1(\R^3)
\] 
and
\be\label{intzero} 
\int_{\R^3}u^*\left(d\langle f,\,y^{-1}\,dy\wedge
y^{-1}\,dy\rangle \right)\,=\,0\,. 
\ee 
Furthermore, for any two such
maps $u$ and $v$ there is a map $\lambda:\R^3\to S^1$ with
$d\lambda\in L^2(\R^3,S^1)$ so that $v=\lambda\, u$. One can choose
$u$ as above with 
\be\label{gaugecond2}
\delta\langle
u^{-1}du,\vp\rangle=0,\quad \text{and} \quad d\langle
u^{-1}du,\vp\rangle\in L^2
\ee 
where $\delta$ is the codifferential.
Such a lift is unique up to left multiplication by a unit complex
number.
\end{proposition}

\smallskip\noindent{\bf Proof.}
The proof is of course similar to the proof of Proposition \ref{rep}. 
We therefore point out the modifications that
are necessary to the previous proof. The first change comes about
when looking for the $1$-form $\xi$. 
Recall that $\xi$ is defined as a solution 
of the elliptic system 
\be\label{dxi}
d\xi = -\frac14 \,\varphi\, d\varphi\wedge
d\varphi\,, \qquad
\delta\,\xi = 0.
\ee
The finite energy condition
tells us that $ -\frac14 \,\varphi\, d\varphi\wedge d\varphi$ is in
$L^2$. We also know that this form  is closed 
(in the sense of distributions). It follows from the Hodge decomposition, Lemma \ref{hodge}, that 
system (\ref{dxi}) has a unique solution $\xi = \xi_i\,dx^i$ with the following properties: first derivatives of each coefficient $\xi_i(x)$ are square-integrable, and each $\xi_i$ is in $L^6$. 
This means that
$a=-\frac12\vp\,d\vp+\vp\xi$ will no longer be in $L^2$, it will be in $L^2 + L^6$.  However,
$a$ is in $L^2$ on any cube so
we can find a sequence of lifts $u_n$ defined on $[-n,n]^3$. The
uniqueness assertion from the compact version of the theorem tells
us that $u_n$ and $u_1$ agree on the unit cube up to a factor in $S^1$.
Multiplying by this factor if necessary the uniqueness result tells
us that the $u_n$ agree on their domains. It follows that we can
define a lift $u$ on  all of $\R^3$.

We now just need to establish equations (\ref{u*d=du*2}),
(\ref{intzero}) and that $u^{-1}du\wedge u^{-1}du \wedge u^{-1}du\in
L^1$. The argument leading to 
\be\label{eqqq} 
u^*\left(d\langle f,
y^{-1}dy\wedge y^{-1}dy \rangle\right) =
\frac13\;u^*(X_m(f^m))\langle a,\,a\wedge a \rangle\,, 
\ee
works without modification.  Turning to the left hand side of
equation (\ref{u*d=du*2}) we can only approximate $\xi$ by a
sequence $\xi_n$ of smooth compactly supported forms such that
$d\xi_n\to d\xi$ in $L^2$ and $\xi_n\to \xi$ in $L^6$.  We set
\[
a_n := -\frac12\,\vp\,d\vp+\vp\,\xi_n\,,
\]
obtaining $a_n\wedge a_n = \frac14\;d\vp\wedge d\vp-d\vp\wedge\xi_n$ and
notice that 
\[
d\langle u^*f, a_n\wedge a_n\rangle \to d\langle u^*f, a\wedge
a\rangle\quad\hbox{as distributions.}
\]
Also because it is a local statement 
(i.e., we can fix a test function and then work in a compact set containing
the support of that test function and apply the argument from the
compact case),   
\[
d\langle u^*f, a_n\wedge a_n\rangle \to u^*\left(d\langle f,
y^{-1}dy\wedge y^{-1}dy \rangle\right)\quad\hbox{as distributions}.
\]
To establish that $u^{-1}du\wedge u^{-1}du \wedge u^{-1}du\in
L^{3/2}$ we cannot restrict to a compact set, but the decomposition
$\langle a,\,a\wedge a\rangle=-\frac34\;\xi\wedge\vp\,d\vp \wedge d\vp$
still holds and this is good enough as $\xi\in L^6$, $\vp$ is
bounded and $d\vp\wedge d\vp\in L^2$. To see that $u^{-1}du\wedge
u^{-1}du \wedge u^{-1}du\in L^{1}$, notice that $d\vp\in L^2$
implies that $d\vp\wedge d\vp\in L^1$ so $d\vp\wedge d\vp\in
L^{6/5}$ by interpolation. Combined with the decomposition of
$\langle a,\,a\wedge a\rangle$, this gives the result.

Notice that together with equation (\ref{eqqq}) this implies that
$u^*\left(d\langle f, y^{-1}dy\wedge y^{-1}dy \rangle\right)$ is in
$L^1$. Now $a\wedge a=\frac14d\vp\wedge d\vp-d\vp\wedge\xi$ is in
$L^{3/2}$ as $d\vp\in L^2$, $\xi\in L^6$ and the first term is in
both $L^1$ and $L^2$. Working in spherical coordinates
$(r,\vartheta)$ this implies that there is a sequence $R_n\to\infty$
such that
\[
\int_{S^2}|\langle u^*f,a\wedge
a\rangle(R_n,\vartheta)|^{3/2}d\,\text{vol}_{S^2}\le 1/(R_n^3\ln
R_n).
\]
We now have
\begin{align*}
|\int_{\R^3}u^*\left(d\langle f, y^{-1}dy\wedge y^{-1}dy
\rangle\right)|&=|\lim_{n\to\infty}\int_{D_{R_n}}d\langle
u^*f,a\wedge a\rangle|\\
&=|\lim_{n\to\infty}\int_{\partial D_{R_n}}\langle
u^*f,a\wedge a\rangle|\\
&\le\lim_{n\to\infty}R_n^2\int_{S^2}|\langle u^*f,a\wedge
a\rangle(R_n,\vartheta)|d\,\text{vol}_{S^2}\\
&\le\lim_{n\to\infty}R_n^2\left(\int_{S^2}|\langle u^*f,a\wedge
a\rangle(R_n,\vartheta)|^{3/2}d\,\text{vol}_{S^2}\right)^{2/3}(4\pi)^{1/3}
=0\,.
\end{align*}
 \hfill$\square$


\section{The primary invariant}\label{sec2}

In this section, we define the pull-back map on second cohomology
for finite Faddeev energy maps. We also prove that two $S^2$-valued
maps with at least this much regularity and the same induced map on
the second cohomology $H^2$ are related by a family of isometries of
$S^2$. This is a generalization of Theorem \ref{our} (\cite[Lemma
1]{AK2}) which proves the same result for smooth maps.


\subsection{Definition}\label{defprim}

The first problem involved in generalizing Pontrjagin's theorem for
Sobolev maps is to identify a suitable cohomology theory where one
can associate an element to each $S^2$-valued map.  While the third
cohomology of any oriented $3$-manifold is isomorphic to the
integers, the second cohomology of an oriented $3$-manifold may have
torsion,  e.g. $H^2({\mathbb R}P^3;\Z)\cong \Z_2$. Thus the deRham
model is no longer sufficient. We use \v{C}ech cohomology. Details
about \v{C}ech theory can be found in \cite{Spanier}. Recall the
natural way to define pull-backs in \v Cech theory. Given a
continuous map $\varphi:X\to Y$ and acyclic open covers ${\mathcal
U}=\{U_\alpha\}_{\alpha\in{\mathcal A}}$ and ${\mathcal
V}=\{V_\beta\}_{\beta\in{\mathcal B}}$ of $X$ and $Y$ respectively
such that for every $U_\alpha\in{\mathcal U}$, there is a
$V_\beta\in{\mathcal V}$ such that $\varphi(U_\alpha)\subseteq
V_\beta$, one can define a pull-back on the \v Cech cohomology.
(Notice that such covers exist for any continuous map, since an
acyclic  cover on $X$ may be refined by intersecting it with the
inverse image of the acyclic cover on $Y$. It is not obvious how to
do the analogous construction with a Sobolev map.) If $\mu$ is the
\v Cech $k$-cycle on $Y$ represented by the collection of integers
$m_{\beta_0\dots\beta_k}$, the pull-back $\varphi^*\mu$ is defined
to be the class represented by  the collection of integers
$n_{\alpha_0\dots\alpha_k}=\sum_{\varphi(U_{\alpha_j})\subseteq
V_{\beta_j}} m_{\beta_0\dots\beta_k}$.

As we now show, we can fix a cover on $M$ and define a class
$\varphi^*\mu_{S^2}\in \check H^2(M;\Z)$ for every finite Faddeev energy map $\varphi$ (without changing the cover) such that it will agree with
the usual notion for smooth maps $\varphi$. The idea here is to
construct a cover by {cubes with cubic intersections} and apply the
local lifting proposition from the previous section to the map
restricted to each cube. The lifts over the various cubes cannot
agree on all of the overlaps (unless $\vp^*\mu_{S^2}=0$) so there
must be circle-valued maps relating the lifts. These circle-valued
lifts form a \v{C}ech cocycle representing a class in $\check
H^1(M,S^1)$. Finally, we use the isomorphism of the first cohomology
with coefficients in the multiplicative group $S^1$ with the second
cohomology with integer coefficients, $H^1(M;\,S^1)\cong
H^2(M;\,\mathbb Z)$, to obtain our definition of the class
$\vp^*\mu_{S^2}$.


For the constructions below we need a triangulation of our
$3$-manifold $M$ such that any nonempty intersection of closed stars
of vertices is piecewise smoothly equivalent to the unit cube in
$\,\mathbb R^3$. We used such a triangulation in \cite{AK} as well.
See Remark \ref{triang} for further discussion.
\begin{definition}\label{cubelike}
A triangulation of a manifold is called cube-like if any nonempty
intersection of closed stars of vertices is piecewise smoothly
equivalent to the unit cube in $\,\mathbb R^n$.
\end{definition}
Fix a cube-like triangulation. Let $U_p$ denote the open star of the
vertex $p$, $U_{pq}$ denote the intersection $U_p\cap U_q$, $U_{pqr}
= U_p\cap U_q\cap U_r$, etc.

Given  a finite energy map $\varphi:M\to S^2$, for every $U_p$ pick
a local lift $u_p:\,U_p\to S^3$ so that $\vp  = u_p^{-1}{\bf
i}\,u_p$ and $u_p\in W^{1,2}(U_p;\,S^3)$. That such a lift exists is
guaranteed by Proposition \ref{rep}. Set
$\lambda_{pq}=u_p\,u_q^{-1}$ on $U_{pq}$. By Proposition \ref{rep},
this is an $S^1$-valued map and $\lambda_{pq}\in
W^{1,2}(U_{pq};\,S^1)$. Clearly, $\lambda_{qp} = \bar \lambda_{pq} =  \lambda_{pq}^{-1}$.
On the non-empty triple intersections $U_{pqr}$ the functions
$\lambda_{pq}$ satisfy the cocycle condition \be\label{lam-co}
\lambda_{qr}\,\lambda_{rp}\,\lambda_{pq}\, =\, 1\,. \ee The maps
$\lambda_{pq}$ can be written in the form
\[
\lambda_{pq}(x)\,=\,e^{2\pi\,{\bf i}\,\theta_{pq}(x)}\,,
\]
for some real-valued functions $\theta_{pq}\in
W^{1,2}(U_{pq};\,\mathbb R)$. This follows from \cite[Lemma 1]{BZ},
or from our Lemma \ref{poi} (after noticing that the ${\bf
i}\,\mathbb R$-valued $1$-form $\lambda_{pq}^{-1}\,d\lambda_{pq}$ is
in $L^2$). Define the function \be\label{n} n_{pqr}\, =\,
\theta_{qr}\,+\,\theta_{rp}\,+\,\theta_{pq} \ee on $U_{pqr}$. In
view of (\ref{lam-co}), $\exp({2\pi {\bf i}\,n_{pqr})} = 1$, so
$n_{pqr}$ must take integer values. If the map $\varphi$ were
smooth, it would follow that the functions $n_{pqr}$ were smooth and
hence constant (as is any mapping from a connected set into a
discrete space). With Sobolev maps the situation is more subtle
since such maps can be discontinuous. However, our functions are in
$W^{1,1}$ and this is enough. Indeed the following result allows one
to conclude that the image of a subset of full measure in $U_{pqr}$
is connected.
\begin{proposition}\label{aeconn} If $X$ is a smooth, compact, connected
manifold, $Y$ is a compact subset of $\R^N$, and $u\in W^{1,1}(X,Y)$, then
there is a connected component $Y_u$ of\  $Y$ so that $u(x)\in Y_u$ for almost every $x$.
\end{proposition}
This result is proved in Appendix \ref{ap2}. Alternatively, we could
have quoted a result of Giaquinta-Modica-Sou\v{c}ek \cite{GMS} that
is sufficient. Thus, there is a fixed integer that we will also
denote $n_{pqr}$ such that $n_{pqr}(x)=n_{pqr}$ for almost all
$\,x\in \overline U_{pqr}$. It is not hard to see that
\begin{equation}\label{cocycle}
n_{pqr}- n_{\ell qr} + n_{\ell pr} - n_{\ell pq} = 0\,,
\end{equation}
provided $U_{\ell pqr}\neq\emptyset$. Thus, $n_{pqr}$ defines a
\v Cech 2-cocycle in the constant sheaf $\mathbb Z$.

\begin{definition}\label{localall} Given $\vp\in W^{1,2}(M,S^2)$, we call
$u_p$, $\lambda_{pq}$, $\theta_{pq}$, and $n_{pqr}$ its local
representatives, transition functions, lifted transition functions,
and cocycle, respectively. The primary homotopy invariant is the
cohomology class represented by $n_{pqr}$ and it is denoted by
$\vp^*\mu_{S^2}$.
\end{definition}

\begin{remark}
Notice that this defines the pull-back for fairly singular functions
such as those from Example 5.  This approach to define
$\vp^*\mu_{S^2}$ would not work for $\vp\in W^{1,p}(M,S^2)$ for
$p<3$ without additional regularity assumptions. Indeed, it is
unlikely that there is any reasonable definition of the pull-back
for such functions in light of the function $\phi:D^3\to S^2$ from
Example 4 given by $\phi(x)=x/|x|$.
\end{remark}

Since the local representatives are not unique, we should explain
why $\vp^*\mu_{S^2}$ is well-defined. By Proposition \ref{rep}, any
other set of representatives is obtained by multiplying each $u_p$
on the left by some $S^1$-valued function $\mu_p\in W^{1,2}$. We
thus write another set of representatives, $\tilde u_p =
\mu_p\,u_p$, and another set of transition functions,
$\tilde\lambda_{pq} = \tilde u_p\,\tilde u_q^{-1}$. The transition
functions $\tilde\lambda_{pq}$ are related to the transition
functions $\lambda_{pq}$ by the equation
\[
\lambda_{pq}\;\tilde\lambda_{pq}^{-1}\,=\,\mu_p^{-1}\,\mu_q\,,
\]
meaning, of course, that $\lambda_{pq}\;\tilde\lambda_{pq}^{-1}$ is
a coboundary. The choice of the lifts $\theta_{pq}$ also fails to be
unique: we can replace $\theta_{pq}(x)$ by ${\widehat\theta}_{pq}(x)
= \theta_{pq}(x) + \gamma_{pq}$ with integer $\gamma_{pq}$. Defining
$\xi_p$ by $\mu_p^{-1}\,d\mu_p = 2\pi{\bf i}\,d\xi_p$, and
incorporating $\gamma_{pq}$, we obtain the equation relating the
lifted transition functions: $\tilde\theta_{pq}(x)  = \theta_{pq}(x)
+  \gamma_{pq} + \xi_p(x) - \xi_q(x)$. The effect this makes on the
cocycle $n_{pqr}$ is this:
$$
{\tilde n}_{pqr} = n_{pqr} + \gamma_{pq} + \gamma_{qr} + \gamma_{rp}\,.
$$
This equation represents the fact that
${\tilde n}_{pqr}$ and $n_{pqr}$ belong to the same
$2$-cohomology class.

This extension of the pull-back of the fundamental class to Sobolev
maps satisfies everything that one could hope for.  The next
proposition verifies that it coincides with the usual definition for
smooth maps and that every cohomology class is represented as the
primary invariant of some smooth map. The proposition afterward
extends an important lifting result from the smooth case.


\begin{proposition}\label{eqsmooth}
The  class $\varphi^*\mu_{S^2}$ given in Definition \ref{localall} exactly
coincides with the usual definition for smooth maps $\varphi$.
Furthermore, given any Sobolev map for which $\varphi^*\mu_{S^2}$ is
defined, there is a smooth map with the same value.
\end{proposition}

\smallskip\noindent{\bf Proof.}
We first identify the associated cohomology class in the smooth
case. We have $\varphi=u_p^{-1}{\bf i}\,u_p$. In the smooth case the
$u_p$'s are sections of the bundle $Q_{\varphi{\bf i}}=\{(x,q)\in
M\times S^3 \,|\; {\bf i}=q\,\varphi(x)\,q^{-1} \}$. It follows that
$\lambda_{pq} = u_p\,u_q^{-1}$ are the transition functions of the
complex line bundle associated with the  principal bundle
$Q_{\varphi{\bf i}}$. The cocycle $n_{pqr}$ represents (minus) the
first Chern class of the line bundle associated to $Q_{\varphi{\bf
i}}$, \cite{GH}. By this and the proof of \cite[Lemma 1]{AK2}, we
have
$$ [n_{pqr}]=c_1(Q_{\varphi{\bf i}})=\varphi^*\mu_{S^2}-{\bf
i}^*\mu_{S^2}=\varphi^*\mu_{S^2}\,.
$$
It turns out that every second cohomology class $\beta \in
H^2(M;\Z)$ is represented as $\varphi^*\mu_{S^2}$ for some smooth
map $\varphi:\,M\to S^2$. We give two arguments for this, one that
quotes well-known results from topology and one that follows from
the definition. The first argument uses that $H^2(M;\Z)$ is in
one-to-one correspondence with the homotopy classes of maps from $M$
to the Eilenberg-MacLane space $K(\mathbb Z, 2)$ which is ${\mathbb
C}P^\infty$. Now the homotopy classes of maps into ${\mathbb
C}P^\infty$ is the same as the homotopy classes of maps into
${\mathbb C}P^2$ by general position. It follows that any given
class in $H^2(M;\Z)$, may be represented as the pull-back of a map
$f:M\to\CP^2$. By general position this map may be assumed to miss
$[0:0:1]$. Composing with the natural projection
$\CP^2-\{[0:0:1]\}\to S^2$ gives the desired map.

The second argument is more direct. Let $\alpha\in H^2(M;\Z)$ be
given. All we need is to construct a continuous map $\tilde\vp$ so
that $\tilde\varphi^*\mu_{S^2}=\alpha$, because this $\tilde\vp$ can
be continuously deformed into a smooth map $\vp$, and the cohomology
class  will not change under homotopy. The continuous map is defined
on the triangulated $M$ as follows. First, map the whole
$1$-skeleton into a fixed point $p\in S^2$. Next, define a map on
$2$-simplices. Pick a functional on $2$-chains $n$ representing
$\alpha$. For each $2$-simplex  $\sigma^2$, take any degree
$n(\sigma^2)$ map from $\sigma^2/\partial\sigma^2$ into $S^2$ with
the boundary, $\partial\sigma^2$, going into $p$. This defines a map
on the $2$-skeleton. The resulting map can be extended to
$3$-simplices $\sigma^3$ because $n$ is closed, i.e.,
$n(\partial\sigma^3) = 0$ for any $\sigma^3$. (This is just the
$2$-cocycle condition (\ref{cocycle}).) Thus, we obtain $\tilde\vp$.
\hfill $\square$


\subsection{Global intertwining maps}\label{g-lift}

We can now prove a global lifting result using the primary homotopy
invariant. It is a generalization of \cite[Lemma 1]{AK2} to Sobolev
maps.

\begin{theorem}\label{lift}
For two finite energy maps $\varphi,\;\psi:M\to S^2$ to be
intertwined, \be\label{intertw} \varphi(x) = \Phi(x)\,\psi(x)\,
\Phi(x)^{-1}\,, \ee by a map $\Phi\in W^{1,2}(M,S^3)$ it is
necessary and sufficient that $\varphi^*\mu_{S^2}=\psi^*\mu_{S^2}$.

If $\varphi^*\mu_{S^2}=\psi^*\mu_{S^2}$ \textbf{ and in addition
$\psi$ is smooth}, then there exists a {\bf cartesian} intertwining
map $\Phi\in W^{1,2}(M,S^3)$, i.e., in addition to (\ref{intertw}),
$\hbox{\rm Re}\left(\Phi^{-1}d\Phi\wedge \Phi^{-1}d\Phi\wedge
\Phi^{-1}d\Phi\right) \in L^1(M)$ (it is in fact in $L^{3/2}$) and,
for any smooth $\R^3$-valued function $f$ on $S^3$,
\be\label{w*d=dw*2} \Phi^*\left(d\langle f,\,y^{-1}\,dy\wedge
y^{-1}\,dy\rangle \right)\,=\, d\left(\Phi^*\left(\langle
f,\,y^{-1}\,dy\wedge y^{-1}\,dy\rangle \right)\right)\,, \ee on $M$
in the sense of distributions. Furthermore $(\Phi^{-1}d\Phi)^{\wedge
2}$ is in $L^{3/2}$ and $\delta\langle\Phi^{-1}d\Phi,\psi\rangle=0$.

If $\Phi_1$ and $\Phi_2$ are two different intertwining maps, then
there is a $\lambda$ in $W^{1,2}(M,S^1)$ such that
$\Phi_1(x){\mathfrak q}(\psi(x),\lambda(x))\,=\,\Phi_2(x)$, where
${\mathfrak q}$ is the map from $S^2\times S^1$ to $S^3$ defined by
the formula \be\label{qfrak} \mathfrak q(x, \lambda) =
q^{-1}\,\lambda\,q,\quad\hbox{\rm for any}\quad q\in
S^3\quad\hbox{\rm such that}\;\quad x = q^{-1}{\bf i}\,q\,. \ee If
$\delta\langle\Phi_k^{-1}d\Phi_k,\psi\rangle=0$ for $k=1$, $2$ then
$\langle\lambda^{-1}d\lambda,{\bf i}\rangle$ is harmonic and
$\lambda$ is smooth.
\end{theorem}

\noindent{\bf Proof.}  We work with a cube-like triangulation $K$.
First assume that there is a map $\Phi$ with $\varphi=\Phi\,\psi\,
\Phi^{-1}$. Let $v_p$ be the local representatives for $\psi$. It
follows that $u_p=v_p\Phi^{-1}$ are local representatives for
$\varphi$. The corresponding transition functions
$\lambda_{pq}=v_p\,\Phi^{-1}\,\Phi\,v_q^{-1}$ agree identically with
the transition functions for $\psi$. It follows that
$\psi^*\mu_{S^2} = \varphi^*\mu_{S^2}$.

 Now assume that
$\psi^*\mu_{S^2} = \varphi^*\mu_{S^2}$ and construct an intertwining
map $\Phi$. We patch it together from local pieces. Let $u_p$,
$\lambda_{pq}$, $\theta_{pq}$, $n_{pqr}$ ($v_p$, $\kappa_{pq}$,
$\vartheta_{pq}$, $m_{pqr}$) be the the  local representatives,
transition functions, lifted transition functions and cocycle
corresponding to $\varphi$ ($\psi$ respectively). Recall that this
means that
\begin{align*}
\vp & = u_p^{-1}\,{\bf i}\,u_p, \\
\lambda_{pq}&=u_p\,u_q^{-1} = 
\exp\left(2\pi\,{\bf i}\,\theta_{pq}\right), \\
 n_{pqr}&=\,
\theta_{qr}\,+\,\theta_{rp}\,+\,\theta_{pq} \quad\text{a.e.},
\end{align*}
and similarly for $\psi$. Also recall from Proposition \ref{rep}
that we can take $u_p\in W^{1,2}$ satisfying the gauge-fixing condition
\[
\delta\langle u_p^{-1}du_p,\vp\rangle=0,\quad \text{and} \quad
i^**(\langle u_p^{-1}du_p,\vp\rangle)=0\,.
\]
In addition, the real $1$-form 
$\xi_p = \langle u_p^{-1}du_p,\vp\rangle$ is the solution of the problem 
\[
d\xi_p = -\frac14\;\vp\, d\vp\wedge d\vp\,,\quad 
\delta\xi_p = 0\,, 
\]
with the boundary condition $i^**\xi_p = 0$. 
Notice that on the intersection $U_{pq}$ the 
difference $\xi_p - \xi_q$ is a harmonic form,
\[
d(\xi_p - \xi_q) = 0,\quad 
\delta(\xi_p - \xi_q) = 0,
\]
and hence, $\xi_p - \xi_q$ is smooth on $U_{pq}$. 
The lifted transition functions $\theta_{pq}$ satisfy the equation
\[
2\pi\,d\theta_{pq}=\langle {\bf
i},\lambda_{pq}^{-1}d\lambda_{pq}\rangle
=\langle\vp,u^{-1}_pdu_p\rangle-
\langle\vp,u^{-1}_qdu_q\rangle=\xi_p-\xi_q\,.
\]
Since $U_{pq}$ is
piecewise smoothly equivalent to a cube, a smooth solution,  $\theta_{pq}$, 
can be written down ``explicitly". In the
appropriate coordinate system (i.e., in the cube, 
with $x_0$ inside the cube) 
\be\label{tpq2}
\theta_{pq}(x)\,= \theta_{pq}(x_0)+\,{1\over 2\pi}\,\int_0^1
\left(\xi_{p} - \xi_q\right)_j(tx)\,x^j\,dt\,. 
\ee
Thus, the functions $\theta_{pq}$ are $C^\infty$. 
Any other solution of the equation
\be\label{thetapq} 
2\pi\,d\theta_{pq} = \xi_p-\xi_q 
\ee
differs from (\ref{tpq2}) by a constant. It should be noted that we cannot claim that 
$\theta_{pq}$ is $C^\infty$ on the closure of 
$U_{pq}$. The best we can say about the behavior 
of $\theta_{pq}$ up to the boundary is that 
$\theta_{pq}\in W^{2,2}(U_{pq})$. Indeed, since 
both $\xi_p$ and $\xi_q$ are in 
$W^{1,2}(U_{pq})$, we get $\theta_{pq}$ in 
$W^{1,2}(U_{pq})$ from (\ref{tpq2}). Differentiating 
equation (\ref{thetapq}) shows that the second derivatives of $\theta_{pq}$ are in $L^2(U_{pq})$. 

In a similar fashion, we have initially 
$\lambda_{pq}\in W^{1,2}(U_{pq})$. However, 
differentiating the equation 
$\lambda_{pq} = \exp\left( 2\pi\,{\bf i}\,\theta_{pq}\right)$ twice (and using the fact that 
the $L^4(U_{pq})$-norm of $d \theta_{pq}$ is bounded 
due to Sobolev imbedding $W^{1,2}\subset L^4$ in 
the three dimensional case), we conclude that 
$\lambda_{pq}\in W^{2,2}(U_{pq})$. In the interior of $U_{pq}$ the function 
$\lambda_{pq}$ is $C^\infty$, of course. 

Let us mention here that, because $U_{pq}$ is bilipschitz equivalent to a cube, its boundary  
satisfies the minimal regularity condition that 
allows us to extend $W^{2,2}$ functions from $U_{pq}$ to the whole manifold. In fact, there exists 
a bounded extension operator 
$E_{pq}:\,W^{2,2}(U_{pq})\to W^{2,2}(M)$, see \cite[Chapter 6]{Stein} and \cite{Adams}. Also, in dimension three the 
$W^{2,2}$ functions are continuous (actually, H\"older continuous). Thus, in particular, 
the functions 
$n_{pqr}(x) = \theta_{pq}(x) + \theta_{qr}(x) + 
\theta_{rp}(x)$ are constant everywhere on $U_{pqr}$.

The same comments hold and we make similar choices for $v_p$, $\kappa_{pq}$,
$\vartheta_{pq}$, $m_{pqr}$.

\bigskip

The definition of the local representatives imply that to have
$\varphi=\Phi\,\psi\, \Phi^{-1}$ locally, we must define $\Phi_p =
u_p^{-1}\,\mu_p\,v_p$, for some $\mu_p:\,U_p=\hbox{st}(p)\to S^1$.
In order for $\Phi_p$ and $\Phi_q$ to agree on $U_{pq}$, we must
have \be\label{cobo} \mu_p \, \mu_q^{-1}
\,=\,\lambda_{pq}\,\kappa_{pq}^{-1}\,. \ee In other words, the maps
$\Phi_p$ agree on the overlaps and define a global map on $M$
exactly when the $1$-cochain $\lambda_{pq}\,\kappa_{pq}^{-1}$ is a
coboundary. That this chain is a coboundary in turn will follow from
our assumption $\psi^*\mu_{S^2}=\varphi^*\mu_{S^2}$. The desired
maps $\mu_p$ are given in equation (\ref{mudef}) below.

By quoting results from algebraic topology, we can arrive at a fast
proof that this chain is a coboundary. However we need a bit more in
order to obtain all of the statements of the theorem. We thus sketch
the argument from algebraic topology and then present the full proof
by unwinding the  quoted results in our particular case.

The short exact sequence of presheaves,
\[
0\to C^0(-,\mathbb{Z} )\to W^{1,2}(-,\R)\to W^{1,2}(-,S^1)\to 0\,,
\]
gives a short exact sequence of chain complexes
\[0\to \check{C}^*(M;{\mathbb Z}) \to \check{C}^*(M; W^{1,2}(-,{\mathbb
R}))\to\check{C}^*(M; W^{1,2}(-,{S^1}))\to 0\,,\] that leads to a
long exact sequence in cohomology. Since the class
$\lambda\kappa^{-1}$ maps to $[n-m]=0$ we see that it must be in the
image of $\check{H}^*(M; W^{1,2}(-,{\mathbb R}))$. Using a partition
of unity we see that $W^{1,2}(-,{\mathbb R})$ is a fine presheaf so
that $\check{H}^*(M; W^{1,2}(-,{\mathbb R}))=0$. It follows that
$\lambda\kappa^{-1}$ is a coboundary. We now spell this out in
greater detail.

The equality of the two classes $\psi^*\mu_{S^2}$,
$\varphi^*\mu_{S^2}$ means that $m_{pqr}$ differs from $n_{pqr}$
only by $\gamma_{qr} + \gamma_{rp} + \gamma_{pq}$, where
$\gamma_{ij}$ are integers. Since \[n_{pqr}\, =\,
\theta_{qr}\,+\,\theta_{rp}\,+\,\theta_{pq}, \] by replacing
$\theta_{pq}$ by $\theta_{pq}+\gamma_{pq}$ we can assume that
$m_{pqr}=n_{pqr}$. This is allowed because $\theta_{pq}$ were only
defined up to an integer choice ($\lambda_{pq}(x)\,=\,e^{2\pi\,{\bf
i}\,\theta_{pq}(x)}$) and the class $\varphi^*\mu_{S^2}$ is
independent of all choices. This implies that
\[
(\theta_{qr} - \vartheta_{qr}) + (\theta_{rp} - \vartheta_{rp}) +
(\theta_{pq} - \vartheta_{pq}) = 0\,.
\]
In other words, the family of real-valued functions $\beta_{pq}(x) =
\theta_{pq}(x) - \vartheta_{pq}(x)$ forms a $1$-cocycle: on
$U_{pqr}$, 
\be\label{beta-co} 
\beta_{qr}(x) + \beta_{rp}(x) +
\beta_{pq}(x) = 0\,. 
\ee 
Up to here we have just reproduced the part
of the argument stating that since the class $\lambda\kappa^{-1}$
maps to $[n-m]=0$  it must be in the image of $\check{H}^*(M;
W^{1,2}(-,{\mathbb R}))$. We even know a bit more: the functions 
$\beta_{pq}$ are smooth by our previous analysis of regularity.

We now turn to the part of the argument that uses the fine
condition. This is an abstract way to say that chains are sums of
chains with small support. Let $\rho_k$ be a partition of unity
subordinate to the cover $\{U_k\}$ and notice that
$\beta_{pq}=\sum_k\rho_k\beta_{pq}$. It is clear that each
$\rho_k\beta$ is coclosed. We will prove that they are all coexact
as well.

We first recall that  each $\beta_{pq}$ 
can be extended to a function 
$\tilde \beta_{pq} = E_{pq}(\beta_{pq})$ 
on the whole manifold $M$ so that 
\[
\|\tilde \beta_{pq}\|_{W^{2,2}(M)}\,\le\,
C\,\|\beta_{pq}\|_{W^{2,2}(U_{pq})}\;. 
\] 
Define functions $\varsigma_p^k$ on $U_p$ of class $W^{2,2}(U_p)$ by
\[
\varsigma_p^k(x):=\left\{\begin{array}{lll}
\rho_k(x)\tilde\beta_{pk}(x) & \text{if}& x\in U_k\\0 &
\text{otherwise}&
\end{array}\right.
\]
To see that \be\label{beta} \rho_k\beta_{pq}(x) = \varsigma_p^k(x) -
\varsigma_q^k(x) \quad\text{on}\quad U_{pq},\ee notice that both
sides are zero if $x$ is not in $U_k$. If $x\in U_k$ then $x\in
U_{pqk}$ so the coclosed condition (\ref{beta-co}) gives
\[
\rho_k\beta_{pq}(x)=\rho_k\beta_{pk}(x)+\rho_k\beta_{kq}(x)=
\varsigma_p^k(x) - \varsigma_q^k(x)\,.
\]
Set
\[
\varsigma_p(x)=\sum_k\varsigma_p^k(x)\,.
\]
To complete the construction of the map $\Phi$, it remains to set
\be\label{mudef}\mu_p = \exp 2\pi{\bf i}\,\varsigma_p\ee and notice
that equation (\ref{cobo}) follows from equation (\ref{beta}) by
exponentiation. The fact that $\Phi\in W^{1,2}(M;\,S^3)$ follows
from local considerations.

\medskip

To prove the next statement of the theorem (the cartesian property), we start by observing
that $\mu_p$ and $\mu_p^{-1}={\mu_p}^*$ both belong to  the Sobolev
space $W^{2,2}(U_p)$. It will be convenient to slightly change our
view of $\mu_p$ by writing $\Phi_p =\Phi|_{U_p}=
\left(\mu_p^{-1}\,u_p\right)^{-1}\,v_p$. We now assume that $\psi$
is smooth and show that the map $\Phi$ we have just constructed is
cartesian. The proof is very similar to the proof in Proposition
\ref{rep}. It will be sufficient to show that each local map
$\Phi_p$ is cartesian. Using $u_p^{-1}du_p=-\vp d\vp+\vp\xi_p$ and
$u_p^{-1}{\bf i}u_p=\vp$, we obtain 
\be\label{xiprime}
\left(\mu_p^{-1}\,u_p\right)^{-1}\,d\left(\mu_p^{-1}\,u_p\right)\,
=\, u_p^{-1}\,du_p\,-\,\vp\,(2\pi\,d\varsigma_p)\,=\,
\frac12\,\vp^{-1}\,d\vp \,+\,\vp\,(\xi_p\,-\,2\pi\,d\varsigma_p).
\ee 
Notice that the real-valued $1$-form
$\xi_p\,-\,2\pi\,d\varsigma_p\in W^{1,2}(U_p)$.

\medskip

We fix a $p$ and simplify the notation writing:
\[
w = \mu_p^{-1}\,u_p\,,\qquad v = v_p\,, \qquad
\Phi\,=\,\Phi_p\,=\,w^{-1}v\,,
\]
and,
\[
\xi\,=\, \xi_p-2\pi d\varsigma_p, \qquad a\,=\,w^{-1}\,dw\,,\qquad
b\,=\,v^{-1}\,dv\,.
\]
Since $\psi$ is smooth, $v$ and $b$ are smooth. On the other hand
$a$ is in $L^2$ and $a\wedge a$ is in $L^{3/2}$. We compute:
\be\label{wdw} \Phi^{-1}\,d\Phi\,=\, b\,-\,\Phi^{-1}\,a\,\Phi\,, \ee
\be\label{wdw^3} (\Phi^{-1}\,d\Phi)^{\wedge 3} \,=\,b^{\wedge
3}-a^{\wedge 3} +\,3\,\hbox{Re}\left( b\wedge b\wedge
\Phi^{-1}\,a\,\Phi - b\wedge \Phi^{-1}a\wedge a\Phi\right) \,. \ee
The same argument used in the proof of Proposition \ref{rep} (namely
decomposing $a$ into components parallel and perpendicular to $\vp$)
shows that $\;a\wedge a \wedge a\;$ belongs to $L^{3/2}(U_p)$. The
form $\;b\wedge b \wedge b\;$ belongs to $L^{3/2}(U_p)$ because it
is the restriction of a smooth function defined on $\overline{U_p}$.
Since $a$ is in $L^2$, $a\wedge a$ is in $L^{3/2}$, $\Phi$ is
bounded and the rest of the factors in the remaining term are smooth
we see that $(\Phi^{-1}\,d\Phi)^{\wedge 3}$ is in $L^{3/2}$ as
claimed.

Given a smooth function $f$ from $S^3$ to $\R^3$ we need to prove
the equality \[ \Phi^*\left(d\langle f,\,y^{-1}\,dy\wedge
y^{-1}\,dy\rangle \right)\,=\, d\left(\Phi^*\left(\langle
f,\,y^{-1}\,dy\wedge y^{-1}\,dy\rangle \right)\right)\,, \] The
proof is a modification of the argument used in Proposition \ref{rep} --
we construct a sequence that approaches both sides of the equation
in the sense of distributions. Recall that on $S^3$,
\[
d\langle f,\,y^{-1}\,dy\wedge y^{-1}\,dy\rangle = \frac13\,
\sum_{i=1}^3X_i(f^i)\, \langle y^{-1}\,dy,\,y^{-1}\,dy\wedge
y^{-1}\,dy\rangle \,,
\]
and hence the left-hand side is
\[
\Phi^*d\langle f,\,y^{-1}\,dy\wedge y^{-1}\,dy\rangle =
\frac13\,\Phi^* \left(\sum_{i=1}^3X_i(f^i)X_i(h^i)\right)\, \langle
c,\,c\wedge c\rangle \,,
\]
where \[c:=\Phi^{-1}d\Phi=b-\Phi^{-1}b\Phi\,.\] Also,
\[
\Phi^*\left(\langle f,\,y^{-1}\,dy\wedge y^{-1}\,dy\rangle \right) =
\langle \Phi^*(f),\,c\wedge c\rangle\,.
\]

To construct our sequence of approximations let $\xi_n$ be a
sequence of smooth $1$-forms converging to $\xi$ in $W^{1,2}$, and
denote \be\label{cn} a_n := \frac12\,\vp^{-1}d\vp + \vp\,\xi_n\,,
\quad c_n:=b-\Phi^{-1}a_n\Phi\,. \ee We have \be\label{c2n}
c_n\wedge c_n:= \frac14 \Phi^{-1}d\vp\wedge d\vp\Phi -
\Phi^{-1}d\vp\wedge \xi_n\Phi-
b\wedge\Phi^{-1}a_n\Phi-\Phi^{-1}a_n\Phi\wedge b+b^{\wedge 2}\,. \ee
Note that
\[
c - c_n\,=\,\Phi^{-1}\vp\,(\xi_n - \xi)\Phi\,,
\]
converges to zero in $W^{1,2}$ and \be\label{c2-} c^{\wedge 2}  -
c^{\wedge 2}_n
=\Phi^{-1}d\vp\wedge(\xi_n-\xi)\Phi+\Phi^{-1}\vp(\xi_n -
\xi)\Phi\wedge b+b\wedge\Phi^{-1}\vp(\xi_n - \xi)\Phi\,, \ee
converges to zero in $L^{3/2}$. This establishes that
$(\Phi^{-1}d\Phi)^{\wedge 2}$ is in $L^{3/2}$. This also implies
that $\langle\Phi^*f,c_n\wedge c_n\rangle\to\langle\Phi^*f,c\wedge
c\rangle$ in $L^{3/2}$, so clearly
\[
d\langle\Phi^*f,c_n\wedge c_n\rangle\to d\langle\Phi^*f,c\wedge
c\rangle\,\quad\text{as distributions.}
\]
By the product rule \be\label{prphf} d\langle\Phi^*f,c_n\wedge
c_n\rangle=\langle d(\Phi^*f),c_n\wedge
c_n\rangle+\langle\Phi^*f,d(c_n\wedge c_n)\rangle \ee We need to
justify this application of the product rule.  As usual we
approximate the various terms. The key to this step is just to
notice that when computing
\[
d(c_n\wedge c_n)=d\left(c_n\wedge c_n-c\wedge c\right)\,,
\]
using equation (\ref{c2-}) the answer tends to zero in $L^1$. To
prove this convergence we use (\ref{c2-}) and compute
\be\label{dc2c2}\begin{aligned} d\left(c_n^{\wedge 2}-c^{\wedge
2}\right)&=d\left(\Phi^{-1}\left(d\vp\wedge(\xi-\xi_n)+\vp(\xi-\xi_n)\wedge\Phi
b\Phi^{-1}+\Phi b\Phi^{-1}\vp(\xi-\xi_n)\right)\Phi\right)\\
&=[c^{\wedge 2}_n-c^{\wedge
2},\Phi^{-1}d\Phi]\\&+\Phi^{-1}d\left(d\vp\wedge(\xi-\xi_n)+\vp(\xi-\xi_n)\wedge\Phi
b\Phi^{-1}+\Phi b\Phi^{-1}\vp(\xi-\xi_n)\right)\Phi\,.
\end{aligned}\ee
We have $\Phi^{-1}d\Phi=b-\Phi^{-1}a\Phi$. Since $b$ is smooth and
bounded and $c^{\wedge 2}_n-c^{\wedge 2}\to 0$ in $L^{3/2}$ the term
$[c^{\wedge 2}_n-c^{\wedge 2},b]\to 0$ in $L^{3/2}$ and hence in
$L^1$. Using equation (\ref{c2-}) and $a=\frac12\vp^{-1}d\vp+\vp\xi$
we also have \[[c^{\wedge 2}_n-c^{\wedge
2},\Phi^{-1}a\Phi]=\Phi^{-1}[d\vp\wedge(\xi-\xi_n)+\vp(\xi-\xi_n)\wedge\Phi
b\Phi^{-1}+\Phi
b\Phi^{-1}\vp(\xi-\xi_n),\frac12\vp^{-1}d\vp+\vp\xi]\Phi\,.\] Since
$\vp$, $b$ and $\Phi$ are bounded and $\xi_n\to\xi$ in $W^{1,2}$
hence in $L^6$, and $d\vp$ is in $L^2$ the only troublesome looking
term in this equation is
$[d\vp\wedge(\xi-\xi_n),\frac12\vp^{-1}d\vp]$. However this last
commutator can be arranged into a bounded factor times $d\vp\wedge
d\vp\wedge(\xi-\xi_n)$ and $d\vp\wedge d\vp$ is in $L^2$ since $\vp$
has finite Faddeev energy. Thus $[c^{\wedge 2}_n-c^{\wedge
2},\Phi^{-1}a\Phi]\to 0$ in $L^{6/5}$ and hence $L^1$. For the last
term in equation (\ref{dc2c2}) notice that $dd\vp=0$, $d\vp$ and
$d\Phi$ are in $L^2$, $db$ is smooth and bounded, and $\xi_n\to \xi$
in $W^{1,2}$ to conclude that the entire last term converges to zero
in $L^1$. This completes the argument showing that $d(c_n\wedge
c_n)\to 0$ in $L^1$ and thus justifying the use of the product rule
in equation (\ref{prphf}). It also shows that the second term of
equation (\ref{prphf}) tends to zero in $L^1$. We now turn to the
first term of (\ref{prphf}).

Using the notation ${\bf e}_1={\bf i}$ etc. 
introduced in the proof of
Proposition \ref{rep} and summation convention, 
we write
\[
d(\Phi^* f) = \Phi^*\left(X_m(f^\ell)\right)\,
\langle c,\,{\bf e}_m\rangle\, {\bf e}_\ell\,.
\]
Thus,
\[
\langle d(\Phi^* f), \,(c\wedge c)_n\rangle \,=\, \Phi^*\left(X_m(f^\ell)\right) \langle
c,\,{\bf e}_m\rangle\wedge \langle (c\wedge c)_n,\,{\bf e}_\ell\rangle
\]
We notice that
\[
\langle c,\,{\bf e}_m\rangle\wedge \langle (c\wedge
c)_n,\,{\bf e}_\ell\rangle= \langle c-c_n,\,{\bf e}_m\rangle\wedge \langle
c_n\wedge c_n,\,{\bf e}_\ell\rangle + \langle c_n,\,{\bf e}_m\rangle\wedge
\langle c_n\wedge c_n,\,{\bf e}_\ell\rangle\,,
\]
and the first term tends to zero in $L^1$. By the general algebraic
computation from the proof of Proposition \ref{rep}, the second term
becomes
\[\frac13\langle c_n,c_n\wedge c_n\rangle\delta_{m\ell}\,.\]
We need to see that this tends to $\frac13\langle c,(c\wedge
c)\rangle\delta_{m\ell}$ in $L^1$. Looking at the expressions for
$c_n$ and $c_n\wedge c_n$ from equations (\ref{cn}) and (\ref{c2n})
we see that the only term that we have to worry about is $\langle
a_n,d\vp\wedge \xi_n\rangle$ because all other terms are products of
factors in $L^2$. For this remaining term we use the formula $a_n =
\frac12\,\vp^{-1}d\vp + \vp\,\xi_n$ to see the magic improvement.
Although $d\vp$ is in $L^2$ and $d\vp\wedge\xi_n$ is in $L^{3/2}$,
the form $d\vp\wedge d\vp\wedge\xi_n$ converges in $L^1$ because
$\vp$ has finite Faddeev energy. This completes the proof of
equation (\ref{w*d=dw*2}).


\medskip
Still assuming that $\psi$ is smooth we now prove that we can pick a
cartesian intertwining map $\Psi$ with
$\delta\langle\Psi^{-1}d\Psi,\psi\rangle=0$. We look for the desired
intertwining map in the form $\Psi=\Phi\,\mathfrak{q}(\psi,\lambda)$.
Recall that the map $\mathfrak{q}$ is defined by
$\mathfrak{q}(\psi,\lambda)=v^{-1}\lambda\, v$ where $\psi=v^{-1}{\bf i}\,v$. 
A direct computation shows that
$\mathfrak{q}(\psi,\lambda)\,\psi\,\mathfrak{q}(\psi,\lambda)^{-1}=\psi$
so that $\Psi$ is also an intertwining map for $\vp$ and $\psi$. We
have
\[
\Psi^{-1} d\Psi = \mathfrak{q}(\psi,\lambda)^{-1} \Phi^{-1} d\Phi\;
\mathfrak{q}(\psi,\lambda) + \mathfrak{q}(\psi,\lambda)^{-1}
d\mathfrak{q}(\psi,\lambda)\,.
\]
Using the invariance of the inner product under conjugation this
gives
\[
\langle \Psi^{-1} d\Psi,\psi\rangle  = \langle  \Phi^{-1}
d\Phi,\psi\rangle + \langle {\mathfrak{q}(\psi,\lambda)}^{-1}
d\mathfrak{q}(\psi,\lambda),\psi\rangle\,.
\]
Now,
\[
{\mathfrak{q}(\psi,\lambda)}^{-1}
d\mathfrak{q}(\psi,\lambda)=v^{-1}dv-\mathfrak{q}(\psi,\lambda)^{-1}v^{-1}dv\;\mathfrak{q}(\psi,\lambda)
+v^{-1} \lambda^{-1} d\lambda\,v\,.
\]
This implies that 
\be\label{qvlam}
\langle\mathfrak{q}(\psi,\lambda)^{-1}d\mathfrak{q}(\psi,\lambda),\psi\rangle=
\langle \lambda^{-1}d\lambda,{\bf i}\rangle\,. 
\ee
Since $\langle\Phi^{-1}d\Phi,\psi\rangle$ is in $W^{1,2}$ by
(\ref{wdw}) and (\ref{xiprime}), we have the Hodge decomposition,
\[
\langle\Phi^{-1}d\Phi,\psi\rangle=d\theta+\delta\alpha+\omega\,,
\]
where $\theta$ is a $W^{2,2}$ function, $\alpha$ is a $W^{2,2}$
$2$-form, and $\omega$ is a harmonic $1$-form.
We take $\lambda=e^{-{\bf i}\theta}$ and compute
\[
\langle\Psi^{-1}d\Psi,\psi\rangle=\langle\Phi^{-1}d\Phi,\psi\rangle+\langle
\lambda^{-1}d\lambda,{\bf i}\rangle =\delta\alpha+\omega\,.
\]
Thus  $\delta\langle\Psi^{-1}d\Psi,\psi\rangle=0$.

We need to check that the new map $\Psi$ is still cartesian. To see
this notice that we can apply the same argument that we used to prove 
that $\Phi$ was cartesian to $\Psi$ the only change is that we will
replace $w$ by $\lambda^{-1}w$.  This changes $\xi$ to
$\xi+d\theta$. This is still in $W^{1,2}$ and this is all we need to
make the argument work.

\medskip

To complete the proof of the theorem, we just need to compare
different intertwining maps. Let $\Phi_1$ and $\Phi_2$ be two
different maps intertwining $\vp$ and $\psi$. Then
$\Phi_1^{-1}\,\vp\, \Phi_1 = \Phi_2^{-1}\,\vp \,\Phi_2$ implies that
$\lambda := u_p\,\Phi_2\Phi_1^{-1}\,u_p^{-1}$ commutes with $\bf i$.
Hence, $\lambda(x)\in S^1$ and $\lambda\in W^{1,2}(M,S^1)$. This
expression for $\lambda$ is independent of the local representative
of $\varphi$ exactly because any other local representative would
have the form $\mu_p u_p$ and both $\mu_p$ and $\lambda$ are
complex. By assumption 
$\psi=\Phi_1^{-1}u_p^{-1}{\bf i}\,u_p\,\Phi_1$,
so
\[
\Phi_1{\mathfrak q}(\psi,\lambda)=\Phi_1\Phi_1^{-1}u_p^{-1}\lambda\; u_p\;\Phi_1
=\Phi_2
\] 
(see (\ref{qfrak})). Now
$d\langle\lambda^{-1}d\lambda,{\bf i}\rangle=0$. Assuming
$\delta\langle\Phi_k^{-1}d\Phi_k,\psi\rangle=0$ for $k=1$, $2$ we
compute
\[
0=\delta\langle\Phi_2^{-1}d\Phi_2,\psi\rangle=\delta\langle\Phi_1^{-1}d\Phi_1,\psi\rangle+\delta\langle
d\mathfrak{q}(\psi,\lambda)\mathfrak{q}(\psi,\lambda)^{-1},\psi\rangle
=\delta\langle\lambda^{-1}d\lambda,{\bf i}\rangle\,,
\]
and conclude that $\langle\lambda^{-1}d\lambda,{\bf i}\rangle$ is
harmonic. It follows that $\lambda$ is smooth. \hfill $\square$

\bigskip


\section{Integrality of the degree}\label{intsec}
The secondary homotopy invariant for a map from a $3$-manifold to
$S^2$ can be interpreted as the degree of a map. In this section we
first discuss integrality results for the degrees of maps in
general, then apply our discussion to finite Skyrme energy maps, and
finally apply the techniques to construct a secondary homotopy
invariant for finite Faddeev energy $S^2$-valued maps. A smooth map
$\Phi:P\to Q$ between closed, connected manifolds of the same
dimension induces a map between the cohomology groups
$\Phi^*:H^{\text{top}}(Q)\to H^{\text{top}}(P)$. As each of these
groups is isomorphic to the integers this map is just multiplication
by some integer. This integer is called the degree of the map. Using
the deRham model one can write
$$
\hbox{deg}(\Phi)=\int_P\Phi^*\omega_Q\,,
$$
where $\omega_Q$ is a normalized volume form on $Q$. This formula
makes sense for sufficiently regular but possibly discontinuous
Sobolev maps. For such maps it is interesting to ask whether the integral is still an integer.
\begin{remark}
The map $\Phi$ from Example \ref{eg5} is the identity on one half of
a sphere and projects points along geodesics from a pole to the
equator on the other half of the sphere. From this description or a
direct computation one can see that
$\int_{S^n}\Phi^*\omega_{S^n}=1/2$ 
(in fact any real number can be achieved
through a map with similar regularity). As we remarked in Example
\ref{eg5}, $\Phi\in W^{1,p}$ for every $p<n$, and  this is not enough
regularity to conclude that the degree is an integer. This example
is close to the borderline of the required regularity.
\end{remark}
\begin{remark}
If $\Psi:P\to Q$ is a map between closed manifolds of the same
dimension for which $\int_P\Psi^*\omega_Q$ is fractional, then $\Psi$
cannot be approximated by smooth functions in any norm strong enough
to imply convergence of the corresponding integrals, because this
integral evaluated for any smooth function is integral. This implies
a close relationship between integrality results and approximation
results.
\end{remark}
Here is a useful general result to prove that the expression for the
degree is an integer (see \cite{LK}).


\begin{proposition}\label{degree}
Let $\Phi:\,P\to Q$ be a $W^{1,1}$ map between two closed manifolds
of the same dimension. Let $\omega_Q$ be a smooth volume form on
$Q$. Assume that $\Phi^*\omega_Q$ is integrable on $P$. In addition,
assume, that $\Phi$ has the following property. For any $h\in
L^\infty(Q)$  such that \be\label{haver}
\int_{Q}\,h\,\omega_{Q}\,\,=\,0\,, \ee the pulled back form
$\Phi^*(h\,\omega_Q)$ is integrable on $P$ and 
\be\label{next}
\int_{ P}\,\Phi^*(h\,\omega_{Q})\,=\,0\,. \ee Then \be \int_P
\Phi^*(\omega_Q)\,=\,\hbox{\rm deg}\,(\Phi)\, \int_Q \omega_Q\,, 
\ee
where $\hbox{\rm deg}(\Phi)$ is an integer. If the map $\Phi$ is
smooth, then $\hbox{\rm deg}(\Phi)$ coincides with the degree of
$\Phi$.
\end{proposition}

\noindent{\bf Proof.} Under our assumptions, there exists an
integrable function $\,N_\Phi( \cdot)\,:\,Q\to {\mathbb Z}\,$ such
that for any scalar $\,h\in L^\infty(Q)$ we have the following {\it
area formula}: \be\label{area} \int_{ P}\,\Phi^*(h\,\omega_{Q})\,
=\,\int_{Q}\,h\,N_\Phi\,\omega_{Q}\,. \ee This formula is justified
using the arguments of \cite[3.2.5, 3.2.20, 3.2.46]{Federer},
\cite[Theorem 2]{Sverak}, \cite[Theorem 6.4]{Muller}. We need to
show that $N_\Phi$ is constant on $Q$. Assume that this function
takes the values $a$ and $b$ on two sets of positive measure, say
$N_\Phi^{-1}(a)$ and $N_\Phi^{-1}(b)$. Now choose
\[
h(x) =(\int\limits_{N_\Phi^{-1}(b)}
\omega_Q\,)\;{\bold 1}_{N_\Phi^{-1}(a)}(x) - (\int\limits_{N_\Phi^{-1}(a)}
\omega_Q\,)\;{\bold 1}_{N_\Phi^{-1}(b)}(x)\,.
\]
This function satisfies equation (\ref{haver}) which, by assumption,
implies (\ref{next}). Thus 
\[ 
(\int\limits_{N_\Phi^{-1}(a)}
\omega_Q\,)(\int\limits_{N_\Phi^{-1}(b)}
\omega_Q\,)(a-b)=\int_{Q}\,h\,N_\Phi\,\omega_{Q}=\int_{
P}\,\Phi^*(h\,\omega_{Q})=0\,.
\]
Thus, $a=b$ and $N_\Phi(x)$ must be constant on $Q$. It is well
known that $N_\Phi$ equals the degree of $\Phi$ for smooth $\Phi$;
see e.g.\cite{Nir}. \hfill$\square$


\begin{remark}\label{plan}
In applications, proving that (\ref{haver}) implies (\ref{next})
amounts to proving that the differential $d$ commutes with the
pull-back  $\Phi^*$. Indeed, equation (\ref{haver}) shows that the
closed form $h\,\omega_Q$ is exact, $h\,\omega_Q = d\alpha$. Hence,
$\int_P \Phi^*(h\,\omega_Q) = \int_P \Phi^*(d\alpha)$, and if
$\Phi^* d = d\Phi^*$, then $\int_P \Phi^*(d\alpha) = \int_P
d(\Phi^*\alpha) = 0$. In general, pull-backs by Sobolev maps (of
relatively low regularity) do not commute with $d$. So, the map,
$\Phi$, should have some special structure or additional
integrability for this to happen. Cartesian maps have the required
additional regularity. This is why we prove that there are cartesian
intertwining maps.
\end{remark}

We could have stated Proposition \ref{degree} using smooth functions
of arbitrarily small support as well because of the following lemma.

\begin{lemma}\label{hsmooth}
If $\Phi\in W^{1,1}(P,Q)$ is such that $\Phi^*\omega_Q\in L^1$ and
$\Phi$ satisfies the implication
$$\int_Q f\omega_Q=0\quad \hbox{implies}\quad
\int_P \Phi^*(f\omega_Q)=0
$$
for smooth functions $f$ of arbitrarily
small support, then it satisfies the implication for all functions
in $L^\infty$.
\end{lemma}

\smallskip\noindent{\bf Proof.}
It is not hard to show that any $L^\infty$ function $h$ on a closed
manifold $Q$ with $\int_{Q} h\,\omega_{Q} = 0$ can be written as a
finite sum of $L^\infty$ functions with arbitrarily small supports
and each of which has zero average as well.  Therefore, from the
very beginning we will assume that the function $h\in L^\infty(Q)$
has conveniently small support on $Q$.

Given $h\in L^\infty(Q)$ with zero average, there exists a sequence
$h_k$ of $C^\infty$ functions with zero average and such that that
they are uniformly bounded and converge to $h$ almost everywhere. To
see this, first mollify $h$ to obtain $C^\infty$ functions $\tilde
h_\epsilon = \int \rho_\epsilon(t-t')h(t') dt'$ (we are in a small
chart, i.e., basically in $\mathbb R^n$). Since $\tilde
h_\epsilon\to h$ in $L^1$, the averages $\langle\tilde
h_\epsilon\rangle:= \int_{Q}\tilde h_\epsilon\,\omega_{Q}$ go to
zero. Pick a smooth function, $\zeta$, with support in the same
chart as $h$ and with average $1$. Now define $h_k = \tilde
h_{\epsilon_k} - \langle\tilde h_{\epsilon_k}\rangle\,\zeta$, where
$\epsilon_k\to 0$.

Now $\int_P \Phi^*(h_k\,\omega_{Q}) =0$ by the assumed implication.
Combining this with the assumption that $\Phi^*\omega_Q\in L^1$ and
the dominated convergence theorem imply that 
\be\label{pullbackzero}
\int\limits_P \Phi^*(h\,\omega_{Q}) = \int\limits_P
h(\Phi)\,\Phi^*\omega_{Q} = 0\,. 
\ee 
\hfill $\square$

The following lemma establishes a special representation of average 
zero $3$-forms on $S^3$ that explains our interest in commuting
pull-back and exterior differentiation applied to forms $\langle
f,y^{-1}dy\wedge y^{-1}dy\rangle$.

\begin{lemma}\label{avzeroform}
Given $h$  a smooth real-valued function on $S^3$ with average $0$:
\[ 
\int\limits_{S^3} h\omega_{S^3}\,=\,0\,,
\]  
there exists a $\R^3$-valued smooth function $f$ on
$S^3$ such that 
\[ 
h\omega_{S^3} \,=\, d\langle f,\,y^{-1}dy\wedge
y^{-1}dy\rangle\,. 
\]
\end{lemma}

\noindent{\bf Proof.} For any scalar function $g$ on $S^3$ we have
\[
dg\,=\,X_1(g)\,\theta^1 + X_2(g)\,\theta^2 + X_3(g)\,\theta^3,
\]
where $\theta=y^{-1}dy$, $X_k$ and $\theta^k$ are the Lie
algebra-valued forms, vector fields and forms introduced in
Proposition \ref{rep}. Thus, for any function $f(y) = {\bold i}
f^1(y) + {\bold j} f^2(y) + {\bold k} f^3(y)$,
\[
d\langle f,\,\theta\wedge \theta\rangle\,=\, \langle
df,\,\theta\wedge \theta\rangle\,=\, \Sigma X_i(f^i)\,\theta^1\wedge
\theta^2\wedge \theta^3\,.
\]
Choose
\[
f^i\,=\,X_i(u)\,,
\]
where $u$ is a solution of the Poisson equation
\[
\,L(u)\,=\,-\frac{1}{2\pi^2}\,h\,.
\]
The corresponding $f$ is the desired function because the
Laplace-Beltrami operator is given by $L:=-\sum_{i=1}^3X_i^2$ and
the volume form is given by
$\omega_{S^3}=\frac{1}{2\pi^2}\theta^1\wedge\theta^2\wedge\theta^3$.
\hfill $\square$


\subsection{Degree for finite Skyrme energy maps}

Our first application of Proposition \ref{degree} shows that the
degree of a finite Skyrme energy map from a closed $3$-manifold into
$S^3$ is integral. Integrality was proved in \cite{EM} for finite
Skyrme energy maps on $\R^3$ by the method outlined in Proposition
\ref{degree} and Remark \ref{plan}. It was also proved for $W^{1,3}$
maps on $\R^3$ by Rivi\`ere, \cite{R2}.

\begin{proposition} \label{S3}
If  the map $\,u\in W^{1,2}(M,\,S^3)\,$ has a finite Skyrme energy
or $\,u\in W^{1,3}(M,\,S^3)\,$ then $ \int\limits_{M}
u^*\omega_{S^3}$ is an integer.
\end{proposition}

\begin{remark}
We define the degree of a finite Skyrme energy map to be the integer
given by this proposition. This extends the usual notion of degree
from smooth maps.
\end{remark}

\smallskip\noindent{\bf Proof.}
Combining Lemma \ref{avzeroform}, Lemma \ref{hsmooth} and
Proposition \ref{degree} we see that it is sufficient to prove that
\[
u^*\left(d\langle f,\,y^{-1}\,dy\wedge y^{-1}\,dy\rangle
\right)\,=\, d\left(u^*\left(\langle f,\,y^{-1}\,dy\wedge
y^{-1}\,dy\rangle \right)\right)\,.
\]
For any finite Skyrme energy map $u:M\to S^3$ both $u^{-1}du$ and
$u^{-1}du\wedge u^{-1}du$ are in $L^2$ and a straightforward
application of the approximation (by mollification, $T_\epsilon$) argument gives:
\[
\begin{split}
d\left(u^*\left(\langle f,\,y^{-1}\,dy\wedge
y^{-1}\,dy\rangle \right)\right) = \lim\limits_{\epsilon\to
0}d\left(\langle T_\epsilon(u^*f),\,T_\epsilon(u^{-1}\,du\wedge
u^{-1}\,du)\rangle
\right)\\
=\lim\limits_{\epsilon\to 0}\left(\langle
dT_\epsilon(u^*f),\,T_\epsilon(u^{-1}\,du\wedge
u^{-1}\,du)\rangle
 +\langle T_\epsilon(u^*f),\,dT_\epsilon(u^{-1}\,du\wedge
u^{-1}\,du)\rangle\right)
\\
= u^*\left(d\langle f,\,y^{-1}\,dy\wedge y^{-1}\,dy\rangle
\right)\,.
\phantom{u^*\left(d\langle f,\,y^{-1}\,dy\wedge y^{-1}\,dy\rangle
\right)}
\end{split}
\]
For a $W^{1,3}$ map, the same approximation argument still works.
\hfill$\square$

The result of the previous proposition generalizes to Chern-Simons
invariants. In this case the set of all possible values is unknown
even in the smooth case (It is conjectured that all such
Chern-Simons invariants are rational.)

\begin{corollary}\label{cs}
If $A$ is a finite Skyrme energy,
$\int_M |A|^2 + |A\wedge A|^2 < \infty$,  distributionally flat $SU(2)$
connection, then there is a smooth connection with the same
holonomy and Chern-Simons invariant.
\end{corollary}

\smallskip\noindent{\bf Proof.}
It is well known that any representation is the holonomy of a smooth
connection. By \cite[Lemma 6]{AK} two flat connections with the same
holonomy are gauge equivalent, so we may assume that our finite
energy flat connection is gauge equivalent to a smooth reference
connection. The Chern-Simons invariants of two gauge equivalent
connections are related by the degree of the gauge transformation.
See the expression above \cite[Lemma 7]{AK}. By Proposition \ref{S3}
this degree is an integer for the finite energy gauge
transformation. Finally we can change the Chern-Simons of the smooth
reference connection by this integer by a suitable smooth gauge
transformation. \hfill$\square$

\begin{remark}
The same results also hold for maps $u:M\to S^3$ and flat
connections in $W^{1,3}$. In the next subsection we address maps to
$S^2$ in the finite Faddeev energy case. All of our results are also
valid for maps in $W^{1,3}$.
\end{remark}

\subsection{The secondary invariant for $S^2$-valued maps}
In this subsection we justify the definition of our numerical
secondary homotopy invariant (\ref{upsilon}) for finite Faddeev
energy maps. Notice that we know that this integral converges by
Theorem \ref{lift}. We will have to prove that this integral is an
integer for sufficiently regular Sobolev maps. This integrality
result follows the general outline described at the start of the
section.

\begin{proposition}\label{upint}
If $\varphi:M\to S^2$ is a finite Faddeev energy map, $\phi:M\to
S^2$ is a smooth map, $\varphi^*\mu_{S^2}=\phi^*\mu_{S^2}$ and
$\Phi:M\to S^3$ is a cartesian map satisfying $\vp=\Phi\phi
\Phi^{-1}$ and $\delta\langle\Phi^{-1}d\Phi,\phi\rangle=0$ as given
by Theorem \ref{lift}, then
$$
\Upsilon(\vp,\phi):=\hbox{\rm
deg}(\Phi):=-\frac{1}{12\pi^2}\int_M\hbox{\rm Re}
\left((\Phi^{-1}d\Phi)^3\right)
$$
is an integer. This is valid for $M$ any closed $3$-manifold or
$\R^3$.
\end{proposition}

\smallskip\noindent{\bf Proof.}
Combining Lemma \ref{avzeroform}, Lemma \ref{hsmooth}, Proposition
\ref{degree} and Theorem \ref{lift} gives the result for closed
manifolds $M$. Using Theorem \ref{R3rep} gives the result for
$\R^3$. \hfill$\square$

We will see that changing the intertwining map only changes
$\Upsilon(\vp,\phi)$ by a multiple of $2m_\vp$ where $m_\vp$ is the
divisibility of $\vp^*\mu_{S^2}$.
\begin{definition}
The divisibility of a class $\beta\in H^2(M;\mathbb{Z})$ is the
unique non-negative integer $m$ such that
\[
(\beta\cup H^1(M;\mathbb{Z}))\cap[M]=m\mathbb{Z}\,.
\]
\end{definition}

This motivates the following definition of a secondary homotopy
invariant for finite Faddeev energy maps.
\begin{definition} Given two finite Faddeev energy maps $\vp$ and
$\psi$ with $\vp^*\mu_{S^2}=\psi^*\mu_{S^2}$ let $m_\vp$ be the
divisibility of the class $\vp^*\mu_{S^2}$. Let $\phi$ be a smooth
map with the same primary invariant as $\vp$ and define the
secondary invariant by
\[
\Upsilon(\vp,\psi)=\Upsilon(\vp,\phi)-\Upsilon(\psi,\phi)\quad(\text{\rm
mod}\;\; 2m_\vp)\,.
\]
For $M=\R^3$ there is no primary invariant so we take
$\psi=\phi={\bf i}$ and write $\Upsilon(\vp)$.
\end{definition}

\begin{remark}
It is obvious that this definition agrees with the numerical
invariant from equation (\ref{upsilon}) on smooth maps -- just take
$\phi=\psi$ and pick the constant map $1$ as the intertwining map
between $\phi$ and $\psi$. The difference in sign from equation
(\ref{upsilon}) is explained by noticing that the result computed
with $\Phi$ is minus the result computed with $\Phi^{-1}$. Comparing
the direction of the intertwining, we see that we have the correct
sign. It is also clear from the definition that
$\Upsilon(\vp,\psi)=-\Upsilon(\psi,\vp)$ and
\[
\Upsilon(\vp,\psi)=\Upsilon(\vp,\chi)+\Upsilon(\chi,\psi)\,.
\]
\end{remark}

The remainder of this section will demonstrate that this secondary
invariant is independent of the choice of smooth map $\phi$ and
gauge-fixed cartesian intertwining maps (Theorem \ref{lift}). The
main ingredient is the following formula for the degree of a product
of two $S^3$-valued functions:
\[
\hbox{\rm deg}\,(\Phi_1\,\Phi_2)\,=\,\hbox{\rm deg}\,(\Phi_1)\,+\,
\hbox{\rm deg}\,(\Phi_2)\,.
\]
This formula follows by a simple application of the product rule to
$\Phi_3=\Phi_1\,\Phi_2$ giving
\[
\hbox{Re}\left((\Phi_3^{-1}d\Phi_3)^{\wedge 3}\right)
\,=\,\hbox{Re}\left((\Phi_1^{-1}d\Phi_1)^{\wedge
3}\right)\,+\,\hbox{Re}\left((\Phi_2^{-1}d\Phi_2)^{\wedge
3}\right)\, -3\,d\;\hbox{Re}\,\left(\,\Phi_1^{-1} d\Phi_1\wedge
d\Phi_2\,\Phi_2^{-1}\,\right)\,.
\]
In concrete applications we just have to make sure that we have
enough regularity to apply the product rule to $\Phi_1\,\Phi_2$ and
$\Phi_1^{-1} d\Phi_1\wedge d\Phi_2\,\Phi_2^{-1}$.

\begin{lemma}\label{phiind}
The secondary invariant $\Upsilon(\vp,\psi)$ is independent of the
choice of intermediate smooth map $\phi$.
\end{lemma}
\noindent{\bf Proof.} Let $\chi$ be a smooth map with the same
primary invariant as $\phi$. By Theorem \ref{our} there is a smooth
intertwining map $\Psi$ so that $\phi=\Psi\chi\Psi^{-1}$. Let $\Phi$
be the intertwining map ($\vp=\Phi\phi\Phi^{-1}$) used to compute
$\Upsilon(\vp,\phi)$. By Theorem \ref{lift} $\Phi$ is in $W^{1,2}$
and $(\Phi^{-1}d\Phi)^{\wedge 2}$ is in $L^{3/2}$.  Clearly
$\Phi\Psi$ is an intertwining map for $\chi$ and $\vp$. That the
product rule is valid on $\Phi\Psi$ follows from the approximation
argument using the fact that $\Phi$ is in $W^{1,2}$. That the
product rule is valid on $\Phi^{-1} d\Phi\wedge d\Psi\,\Psi^{-1}$
follows from the approximation argument using the fact that
$(\Phi^{-1}d\Phi)^{\wedge 2}$ is in $L^{3/2}$ and $\Phi^{-1}d\Phi$
is in $L^{1}$. This implies that
\[
\Upsilon(\vp,\chi)=\Upsilon(\vp,\phi)+\text{deg}(\Psi).
\]
The same formula applies to $\Upsilon(\psi,\phi)$ so the
$\text{deg}(\Psi)$ terms cancel. \hfill $\square$

\begin{remark}
The intermediate smooth map was introduced exactly because we do not
have an argument validating the product rule applied to $\Phi^{-1}
d\Phi\wedge d\Psi\,\Psi^{-1}$ when the two $S^2$-valued maps are
only assumed to have finite energy. We expect that there is a direct
argument validating the product rule but it will require some subtle
cancelation.
\end{remark}

In order to understand the dependence of $\Upsilon(\vp,\psi)$ on the
choice of intertwining maps we recall that any intertwining map  may
be obtained from a fixed one as $\Phi\;{\mathfrak q}(\psi,\lambda)$.
Thus we have to study the degree of the map ${\mathfrak
q}(\psi,\lambda)$. We begin with an elementary calculation of the
degree of the map ${\mathfrak q}:\,S^2\times S^1\to S^3$.
\begin{lemma}\label{qdeg}
The  map ${\mathfrak q}:S^2\times S^1\to S^3$ has degree two.
\end{lemma}

\noindent{\bf Proof.} Recall that $\mathfrak{q}(x,z)=u^{-1}zu$ where
$x=u^{-1}{\bf i}u$. From equation (\ref{a^3}) we have
\[
\mathfrak{q}^*\omega_{S^3}=-\frac{1}{8\pi}x[a,x]^2\frac{1}{2\pi}\langle
a,x\rangle\,,
\]
where
\[
a=\mathfrak{q}^{-1}d\mathfrak{q}=u^{-1}z^{-1}dzu+
u^{-1}du-\mathfrak{q}^{-1}u^{-1}du\mathfrak{q}\,.
\]
The term $u^{-1}z^{-1}dzu$ is parallel to $x$ and the remainder of
$a$ is perpendicular to $x$ so
\[
\frac{1}{2\pi}\langle a,x\rangle=\frac{1}{2\pi{\bf
i}}z^{-1}dz=\omega_{S^1}\,.
\]
Now
\[
x[a,x]^2=x[u^{-1}du,x]^2+x[\mathfrak{q}^{-1}u^{-1}du\mathfrak{q},x]^2
-x[u^{-1}du,x][\mathfrak{q}^{-1}u^{-1}du\mathfrak{q},x]
-x[\mathfrak{q}^{-1}u^{-1}du\mathfrak{q},x][u^{-1}du,x]\,.
\]
Since $\mathfrak{q}$ commutes with $x$ the second term is equal to
the first term. Since $x$ and $u^{-1}du$ anticommute
\[
x[u^{-1}du,x][\mathfrak{q}^{-1}u^{-1}du\mathfrak{q},x]=[u^{-1}du,x]x[\mathfrak{q}^{-1}u^{-1}du\mathfrak{q},x]
=x[\mathfrak{q}^{-1}u^{-1}du\mathfrak{q},x][u^{-1}du,x]\,.
\]
It follows that $x[a,x]^2=2x[u^{-1}du,x]^2=-16\pi\omega_{S^2}$ by
equations (\ref{brasq}) and (\ref{normform}). This completes the
proof. \hfill$\square$

By Theorem \ref{lift} any two gauge fixed cartesian intertwining
maps are related by
\[
\Phi_1=\Phi_2\;\mathfrak{q}(\phi,\lambda)\,,
\]
for some smooth map $\lambda:M\to S^1$. Notice that the computation
in the proof of Lemma \ref{qdeg} is completely algebraic. This
implies that 
\be\label{qpullback}
\left(\mathfrak{q}(\phi,\lambda)\right)^*\omega_{S^3}=2\phi^*\omega_{S^2}\wedge\lambda^*\omega_{S^1}\,.
\ee 
Now consider the map taking $C^\infty(M,S^1)$ to $\R$ given by
\[
\lambda\mapsto\int_M\phi^*\omega_{S^2}\wedge\lambda^*\omega_{S^1}\,.
\]
Since
$(\lambda\mu)^{-1}d(\lambda\mu)=\lambda^{-1}d\lambda+\mu^{-1}d\mu$
this map is a group homomorphism. This is just the degree of the
smooth map $(\phi,\lambda):M\to S^2\times S^1$ which is also
$(\phi^*\mu_{S^2}\cup\lambda^*\mu_{S^1})\cap[M]$. The image of this
map is a subgroup of $\mathbb{Z}$ so it takes the form
$m_\phi\mathbb{Z}$ for some non-negative $m_\phi$. Since every first
cohomology class can be represented by $\lambda^*\mu_{S^1}$ we see
that the number $m_\phi$ is exactly the divisibility of
$\phi^*\mu_{S^2}=\vp^*\mu_{S^2}$.  To finish we just have to verify
that the formula for the degree of the product of $\Phi$ and
$\mathfrak{q}(\phi,\lambda)$ is valid.

\begin{lemma}\label{Phiind}
The secondary invariant $\Upsilon(\vp,\psi)$ is independent of the
choice of intertwining maps.
\end{lemma}

\noindent{\bf Proof.} Since $\Phi$ is in $W^{1,2}$ and
$\mathfrak{q}(\phi,\lambda)$ is smooth and $(\Phi^{-1}d\Phi)^{\wedge
2}$ is in $L^{3/2}$ the standard approximation argument shows that
the product rules for $ d(\Psi\;\mathfrak{q}(\phi,\lambda))$ and
$d(\Phi^{-1}d\Phi\wedge
d\mathfrak{q}(\phi,\lambda)\;\mathfrak{q}(\phi,\lambda)^{-1}$ are
valid so the degree of $\Phi\;\mathfrak{q}(\phi,\lambda)$ is just the
sum of the degrees of $\Phi$ and $\mathfrak{q}(\phi,\lambda)$ as
required. \hfill$\square$

Notice that for $\R^3$ the secondary invariant $\Upsilon(\vp)$ is
well-defined and is an integer. The point is that the intertwining
maps are well-defined up to multiplication by a constant and any
constant map has degree zero. As we saw in Theorem \ref{ourhopf}
when $\vp^*\mu_{S^2}=0$ there is a direct expression for
$\Upsilon(\vp)$ that does not require an intertwining map. The
arguments that we have given combine to show that this is an integer
for finite energy maps as well.

\begin{corollary}
If $\vp:M\to S^2$ has finite Faddeev energy and $M=\R^3$ or
$\vp^*\mu_{S^2}=0$ then there is a $1$-form on $M$, call it
$\theta$, such that $\vp^*\omega_{S^2}=d\theta$ and
\[
\text{\rm Hopf}(\vp) = \int_M\theta\wedge d\theta\,,
\]
is a well-defined integer.
\end{corollary}

\noindent{\bf Proof.} We just take
$\theta=\frac{1}{2\pi}\langle\Phi^{-1}d\Phi,\vp\rangle$ where $\Phi$
is the gauge fixed intertwining map. The remainder of the argument
is just the argument from Theorem \ref{ourhopf} with $N=1$ and
$\alpha=0$.


\section{Conclusion and applications}
We conclude by addressing questions that were posed in \cite{AK},
\cite{AK2} and \cite{K}. The first questions raised in \cite{AK2}
were about the regularity of minimizers of the Faddeev functional.
This is still an interesting open question. The next question was
how to extend obstruction theory for finite energy maps; what
cohomology theory should one use?  how would one define the primary
invariant of a finite energy map? We have given a very satisfactory
answer to these questions in Definition \ref{localall} and Theorem
\ref{lift}. We have defined the pull-back on cohomology for finite
energy maps in a way that generalizes the pull-back for smooth maps.
In addition we showed that the pull-back of the fundamental class on
$S^2$ is the obstruction to the existence of a lift from $S^2$ to
$S^3$ even for finite energy maps. Theorem \ref{lift} generalizes
\cite[Lemma 1]{AK2} from the smooth case as it should. We gave an
example of a discontinuous finite energy map that is not in VMO
(Example \ref{eg5}) so our results are true, but not for trivial
reasons.

Unlike the primary homotopy invariant for maps $\vp:\,M\to S^2$, it
was already clear from \cite{AK2} how to generalize the secondary
invariant for pairs of such maps once the primary invariant was
defined. Likewise it was clear from \cite{AK} how to define the
primary invariant for maps $u:M\to S^3$ or the Chern-Simons
invariants of $SU(2)$-connections. Each of these could be expressed
as an integral and the integral could, in principal converge for
possibly discontinuous finite energy maps. What was not clear is if
the values taken by these integrals for finite-energy maps would
coincide with the values taken by smooth maps. This left open the
possibility of phantom sectors of finite energy maps containing no
smooth maps. Proposition \ref{S3} and Corollary \ref{cs} demonstrate
that this does not occur for $S^3$. The generalization to other Lie
groups is still an interesting open question.

Whereas the integrality of the Hopf invariant is clear for smooth
maps, it was an open question for finite energy maps. The Hopf
invariant is just a special case of the secondary invariant
appearing in Pontrjagin's theorem. Proposition \ref{upint}
establishes that the integral for the secondary invariant does
indeed give an integer. Lemma \ref{phiind}, Lemma \ref{Phiind} and
Lemma \ref{qdeg} show that the integral for the secondary invariant
descends to give a well-defined invariant in the appropriate cyclic
group for finite energy maps.

All of the results that we have discussed thus far would follow
easily from an approximation theorem proving that any finite energy
map could be approximated by smooth maps in a reasonable sense.
Unfortunately this is not known.  Smooth approximation  remains an
interesting question. All of the examples of functions that are not
reasonably approximated by smooth functions that we know arise
because they do not respect homotopy properties satisfied by smooth
maps (for example have fractional degree). We have seen that finite
energy maps satisfy the properties that one would expect based on
the homotopy theory of smooth maps, so we expect that there is a
reasonable way to approximate finite Skyrme and Faddeev energy maps.

In \cite{AK2}, we proved the existence of minimizers of the Faddeev
functional in a slightly bizarre looking function space. The results
of this paper show that the function space was just the space of
finite energy Sobolev maps in a fixed homotopy class.

\begin{proposition}
Let $M$ be a closed orientable $3$-manifold. For any finite energy
$\varphi:M\to S^2$, the functional
$$
E(\psi)\,=\,\int_M |d\psi|^2\,+\,|d\psi\wedge d\psi|^2\,d\hbox{vol}_M\,.
$$
has a minimizer in the class
$$
\{\psi\in W^{1,2}(M,S^2)|E(\psi)<\infty, \ \varphi^*\mu_{S^2}=\psi^*\mu_{S^2}, \
 \Upsilon(\varphi,\psi)=0\;(\text{\rm mod}\; 2m_\vp)\}\,.
$$
Furthermore this class contains a smooth map.
\end{proposition}

\noindent{\bf Proof.} Given a finite energy map $\vp$ we know from
Theorem \ref{lift} that there is a smooth map $\phi$ with the same
primary invariant. By Proposition \ref{upint} we know that
$\Upsilon(\varphi,\phi)$ is an integer. We also know that every
integer is the degree of some smooth map from $M$ to $S^3$ (just
compose the map obtained by collapsing the two-skeleton with a
degree $n$ self map of $S^3$). Let $\Phi$ be a map of degree
$\Upsilon(\varphi,\phi)$ so that there is a smooth map
$\chi:=\Phi\phi\Phi^{-1}$ in the class with
\[\Upsilon(\varphi,\chi)=\Upsilon(\varphi,\phi)-\text{deg}(\Phi)=0\,.\]
 By the main existence result
from \cite{AK2} we know that there is a minimizer of $E$ in the
class of functions denoted by $\mathfrak{A}_\chi$ in \cite{AK2}.
This class  of functions consists of functions
$\psi=\Psi^{-1}\chi\Psi$ such that
\begin{enumerate}
\item $E(\psi)<\infty$
\item $\Upsilon(\psi,\chi)=0 \; (\text{mod}\;2m_\vp)$
\item $\delta\langle\Psi^{-1}d\Psi,\chi\rangle=0$
\item
$\mathcal{H}\langle\Psi^{-1}d\Psi,\chi\rangle=\sum_kh_k\eta_k$,
\end{enumerate}
where $\mathcal{H}$ is the harmonic projection of the form, $h_k\in
[0,1]$ and $\eta_k$ are an integral basis for the image of
$H^1(M;\mathbb{Z})$ under $\mathcal{H}$ in the space of harmonic
$1$-forms. We will see that these two classes of functions coincide.
Indeed, Theorem \ref{lift} tells us that $\phi$ and $\vp$ are
related by an intertwining map in $W^{1,2}$ since they have the same
primary invariant, hence $\vp$ and $\psi$ are related by an
intertwining map obtained as the product of the three intertwining
maps. This implies that any function in $\mathfrak{A}_\chi$ has the
same primary invariant as $\vp$ by Theorem \ref{lift}. The finite
energy condition is given by item 1 and the condition on the
secondary invariant is given by item 2. Working in the other
direction, we see that conditions 1, 2 and 3 hold from the
assumptions on the class described in this theorem together with the
construction of $\chi$ provided that a gauge-fixed intertwining map
is chosen. To obtain the fourth condition just notice that changing
the intertwining map by right multiplication by
$\mathfrak{q}(\chi,\lambda)$ with harmonic $\lambda$ preserves the
first three conditions and changes the value of
$\mathcal{H}\langle\Psi^{-1}d\Psi,\chi\rangle$ by an arbitrary
integral harmonic form. \hfill$\square$

\begin{remark}
The assumption that $M$ is orientable is really not necessary. We
could look for $\Z_2$-equivariant minimizers on the orientation
cover. In this case there would be no secondary invariant.
\end{remark}

Similar questions about classes of functions are raised by S.
Koshkin in \cite{K}. His paper addresses minimization of the Faddeev
functional for maps into homogeneous and symmetric spaces. In \cite{K} the class of admissible maps $\mathcal{E}$ is introduced to
study the minimization problem in a fixed $2$-homotopy sector. The
class $\mathcal{E}$ is defined as the maps $\psi=u^{-1}\vp u$ where
$\vp$ is smooth and $a^\perp\in L^2$, $a^\perp\wedge a^\perp\in L^2$
and $a^{||}\in W^{1,2}$ where $a=u^{-1}du$ and in our case
$a^{||}=\langle a,\vp\rangle\vp$, $a^\perp=\frac12\vp[a,\vp]$.
Koshkin conjectured \cite[Conjecture 1]{K} that the union of the
$\mathcal{E}$ taken over a family of smooth functions $\vp$
representing every homotopy type is exactly the class of finite
energy maps. We can now see that this conjecture is true for
$S^3/S^1$. Indeed let $\psi$ be a finite energy map and let $\vp$ be
a smooth map with the same primary invariant as given by Proposition
\ref{eqsmooth}. Setting $\Phi$ to be the gauge-fixed intertwining
map given by Theorem \ref{lift} we have $\vp=\Phi^{-1}\psi\Phi$ so
\[
a^\perp=\frac12\vp[a,\vp]=\frac12\vp\Phi^{-1}d\psi\Phi-\frac12\vp
d\vp\,,
\]
and
\[
a^{||}=\langle a,\vp\rangle\vp=\langle
u^{-1}du,\vp\rangle\vp-\langle v^{-1}dv,\psi\rangle\vp\,,
\]
where $u$ and $v$ are local lifts of $\vp$ and $\psi$ respectively.
The first term in $a^{||}$ is smooth and with the proper
gauge-fixing the second term will be in $W^{1,2}$. The first term in
$a^\perp$ is in $L^2$ since $\psi$ has finite energy and $\Phi$ is
bounded and the second term is smooth. The only term that one needs
to worry about in $a^\perp\wedge a^\perp$ is the square of the first
term in $a^\perp$. This is just
\[
\frac14\vp\Phi^{-1}d\psi\Phi\vp\Phi^{-1}d\psi\Phi=-\frac14\vp\Phi^{-1}\psi
d\psi\wedge d\psi\Phi\,,
\]
which is also in $L^2$ since $\psi$ has finite energy. Working in
the other direction uses the same formulas to see that $\psi$ has
finite energy given that $a^\perp\in L^2$, $a^\perp\wedge a^\perp\in
L^2$ and $a^{||}\in W^{1,2}$. The argument is contained in
\cite{AK2} and \cite{K}.

Versions of our local representation result (Proposition \ref{rep}),
primary invariant (Definition \ref{localall}) and global
intertwining result (Theorem \ref{lift}) should hold for homogeneous
spaces of the form $G/T^k$ with only minor modifications. This class
includes all flag manifolds. It follows that 
\cite[Conjecture 1]{K}
also holds for these spaces. The definition of the primary invariant
becomes more interesting for homogeneous spaces $G/H$ when $H$ is
nonabelian. One can not take cohomology with values in $H$ as we did
in the proof of Theorem \ref{lift}, but this is probably not needed.
The point is that the primary invariant will arise as a cocycle
analogous to our $n$ but this should be interpreted as the degree of
various overlap maps into $H$.


\section*{Appendix 1: The $W^{1,1}$ image of a connected set}\label{ap2}
\addcontentsline{toc}{section}{Appendix 1: The $W^{1,1}$ image of a
connected set}

The question whether the image of a connected set under a Sobolev map is connected or
not has been discussed in the literature.
For $\,W^{1,1}\,$ maps, there is a
result of Giaquinta, Modica and Sou\v cek, \cite{GMS}, that the image, $\,\bar{u}({\mathcal A}_C(u))$,
\footnote{where
$\,\bar{u}(x)\,=\,\hbox{aplim}_{y\to x} u(y)\,$ for $\,x\in {\mathcal A}_C(u)$}
 of the set of approximate continuity, $\,{\mathcal A}_C(u)$,
 of a map
$\,u\in W^{1,1}(\Omega, \mathbb R^N)\,$ on an open connected set
$\,\Omega\subset\mathbb R^n$,  is connected; see  \cite{GMS} for
details. This would suffice for our purposes, but we give a simple
independent argument proving a convenient substitute connected image
result for Sobolev maps.

\begin{lemma}
Let $\,X\,$ be a smooth, compact, connected manifold of dimension $\,n$, and let $\,Y\,$ be a compact subset of $\,\mathbb R^N$.
For any map $\,f\in W^{s,p}(X;\,Y)$, where $\,s>0$, $\,p\ge 1$, and
$\,s-\frac{n}{p}\,\ge \,1-n$, there exists a connected component, $\,Y_1$, of $\,Y\,$ such that
$\,f(x)\in Y_1\,$ for almost all $\,x\in X$.
\end{lemma}

\noindent{\bf Proof.}\ In view of the embedding $\,W^{s,p}\subset W^{1,1}$, it is
sufficient to consider only the case $\,f\in W^{1,1}$. Using the
description of Sobolev spaces on closed manifolds, choose a finite
open cover $\,\{U_i\}\,$ of $\,X\,$ together with smooth coordinate
maps $\,\chi_i:\,U_i\to 2I^n\,$ with the property that the open sets
$\,\chi_i^{-1}(I^n)\,$ still cover $\,X$. Let $\,\vartheta\,$ be a
smooth cut-off function on $\,X\,$ with support in
$\,\chi_i^{-1}(2I^n)$, which is positive on the set
$\,\chi_i^{-1}(\frac32I^n)\,$ and, in addition, $\,\vartheta(p) =
1\,$ for $\,p\in\chi_i^{-1}(I^n)$. Given $\,f\in W^{s,p}(X;\,Y)$,
the function $\,(\vartheta\cdot f)\circ \chi_i^{-1}\,$ belongs to
$\,W^{s,p}(2I^n; \mathbb R^N)\,$ and has the property that when
restricted to the cube $\,I^n\,$ its values lie in $\,Y$. If we
prove, that, in fact, its values are in one connected component of
$\,Y$, then applying this argument to each chart and using the
connectedness of $\,X\,$ we prove the lemma. Thus, we localize our
consideration to a cube and prove the following result:

\bigskip

\noindent{\it
If
$\,f\in W^{1,1}(I^n(1);\,\mathbb R^N)\,$ is such that $\,f(x)\in Y\,$ for a.e. $\,x\in I^n(1)$, then, except for at most a set of measure $\,0$,
the image of the cube $\,I^n(1)\,$ under the map $\,f\,$
lies entirely in one component of $\,Y$.
}

\bigskip

\noindent The proof is by induction on $\,n$, the
dimension of the cube. If $\,n=1$, the statement of the lemma is
true because any function in $\,W^{1,1}(I^1(1);\,\mathbb R)\,$ is
absolutely continuous after possibly a change on a set of measure
zero (\cite{Ziemer}). Assuming the lemma is true for all maps in
$\,W^{1,1} (I^{n-1}(1);\,\mathbb R^N)$, let us prove it for an
$\,f\in W^{1,1}(I^n(1);\,\mathbb R^N)$. Split the coordinates
$\,x^1,\dots,x^n\,$ into $\,x^\prime = (x^1,\dots, x^{n-1})\,$ and
$\,x^n$. It is not hard to see that (after a possible change on a
set of measure zero in $\,I^n$) $\,f(\cdot, t)\in W^{1,1}
(I^{n-1};\,\mathbb R^N)$, for a.e. $\,t\in I^1$. Moreover, there is
a set $\,\Theta \in I^1\,$ of measure $\,1\,$ such that, for every
$\,t\in \Theta$, not only $\,f(\cdot, t)\in W^{1,1}
(I^{n-1}(1);\,\mathbb R^N)$, but also $\,f(x^\prime, t)\in Y$, for
almost all $\,x^\prime\in I^{n-1}(1)$. On the other hand, it is well
known (see \cite{Ziemer}, 2.1.4) that a function in $\,W^{1,1}\,$ is
absolutely continuous on almost every line segment parallel
to the coordinate axes. 
Thus, there is a set $\,\Xi\in I^{n-1}(1)\,$ of measure $1$ such that
$\,f(x^\prime,\cdot)\,$ is absolutely continuous for all $\,x^\prime\in\Xi$.
At this point we stop adjusting $\,f$.

For every $\,t\in \Theta$, by the induction assumption, there is a
set $\,\Sigma_t\in I^{n-1}(1)\,$ of full measure and a connected
component $\,Y_t\,$ of $\,Y\,$ such that $\,f(x^\prime, t)\in Y_t$,
for all $\,x^\prime\in \Sigma_t$. Assume that for some $\,t_1\,$ and
$\,t_2\,$ the corresponding components $\,Y_{t_1}\,$ and
$\,Y_{t_2}\,$ are different. Take any $\,x^\prime\,$ from the
intersection $\,\Sigma_{t_1}\cap \Sigma_{t_2}\cap\Xi$ (which has
measure $1$). Then $\,f(x^\prime, x^n)$, for all $\,x^n\in I^1$,
should lie in the same component of $\,Y$. A contradiction. This
shows that all $\,Y_t\,$ are the same and proves the lemma.

\begin{remark}
Following Brezis and Nirenberg \cite{BN}, define the {\it essential range},
$\,\hbox{ essRange}(f)$,  of a measurable map $\,f\,$ from a compact set $\,X\subset \mathbb R^n\,$
into a compact set $\,Y\subset\mathbb R^N\,$ as the smallest closed set $\, \mathfrak C\,$ in $\,Y\,$ such that
$\,f(x)\in\mathfrak C\,$ for almost every $\,x\in X$. The essential range is well defined, as shown in \cite{BN}. Proposition \ref{aeconn} shows that
the essential range of a $\,W^{1,1}\,$ map from a connected manifold
into a a compact subset of $\,\mathbb R^N\,$ lies entirely in
one connected component of the target set.
\end{remark}


\end{document}